\documentclass[
amssymb,preprint,preprintnumbers,nofootinbib,superscriptaddress]{revtex4}
\usepackage{amsmath}
\usepackage{graphicx}
\graphicspath{{Figures/}}
\usepackage{latexsym}
\usepackage{amsfonts}
\usepackage{url,hyperref}
\usepackage{bm}
\usepackage{textcomp}
\usepackage{xcolor}
\usepackage{bbm}
\usepackage{slashed}
\usepackage{caption}
\usepackage{epstopdf}
\usepackage{subcaption}
\usepackage{placeins}
\usepackage{color}
\usepackage[a4paper,left=20mm,right=20mm,top=25mm,bottom=25mm]{geometry}

\begin{document}
\title{Thermodynamics of Asymptotically de-Sitter Black Hole in dRGT Massive Gravity from R\'{e}nyi entropy}

\author{Phuwadon Chunaksorn \footnote{Email: maxwelltle@gmail.com}}
\affiliation{The Institute for Fundamental Study, Naresuan University, Phitsanulok, 65000, Thailand}
\affiliation{Thailand Center of Excellence in Physics, Ministry of Higher Education, Science, Research and Innovation, 328 Si Ayutthaya Road, Bangkok 10400, Thailand}

\author{Ekapong Hirunsirisawat \footnote{Email: ekapong.hir@kmutt.ac.th}}
\affiliation{Quantum Computing and Theory Research Centre (QX), Faculty of Science, King Mongkut's University of Technology Thonburi (KMUTT), Pracha Uthit Road, Bangkok, 10140, Thailand}
\affiliation{Learning Institute, King Mongkut's University of Technology Thonburi (KMUTT), Pracha Uthit Road, Bangkok, 10140, Thailand}

\author{\\Ratchaphat Nakarachinda \footnote{Email: tahpahctar\_net@hotmail.com}}
\affiliation{The Institute for Fundamental Study, Naresuan University, Phitsanulok, 65000, Thailand}
		
\author{Lunchakorn Tannukij \footnote{Email: lunchakorn.ta@kmitl.ac.th}}
\affiliation{Department of Physics, School of Science, King Mongkut's Institute of Technology Ladkrabang, 1 Chalong Krung 1 Alley, Lat Krabang, Bangkok, 10520, Thailand}

\author{Pitayuth Wongjun \footnote{Email: pitbaa@gmail.com}}
\affiliation{The Institute for Fundamental Study, Naresuan University, Phitsanulok, 65000, Thailand}
\affiliation{Thailand Center of Excellence in Physics, Ministry of Higher Education, Science, Research and Innovation, 328 Si Ayutthaya Road, Bangkok 10400, Thailand}

\begin{abstract}
 The thermodynamic properties of the de Rham-Gabadadze-Tolley (dRGT) black hole in the asymptotically de Sitter (dS) spacetime are investigated by using R\'enyi entropy. It has been found that the black hole with asymptotically dS spacetime described by the standard Gibbs-Boltzmann statistics cannot be thermodynamically stable. Moreover, there generically exist two horizons corresponding to two thermodynamic systems with different temperatures, leading to a nonequilibrium state.  Therefore, in order to obtain the stable dRGT black hole, we use the alternative R\'enyi statistics to analyze the thermodynamics properties in both the separated system approach and the effective system approach. Interestingly, we found that it is possible concurrently obtain positive pressure and volume for the dRGT black hole while it is not for the Schwarzschild-de Sitter (Sch-dS) black hole. Furthermore, the bounds on the nonextensive parameter for which the black hole being thermodynamically stable are determined.  In addition, the key differences between the systems described by different approaches, e.g., temperature profiles and types of the Hawking-Page phase transition are pointed out. 
\end{abstract}

\maketitle

\section{Introduction}\label{intro}

\par
General relativity (GR) has been verified by several recent astrophysical observations. Nevertheless, the discovery of late time accelerated expansion of the universe \cite{SupernovaSearchTeam:1998fmf,SupernovaCosmologyProject:1998vns} has led to the curiosities, and also doubts, among the community in the nature of gravitation, namely the theory itself, at the cosmic scale. Based on GR, dark energy needs to be proposed in describing the cosmic accelerated expansion. Despite not knowing the true candidate(s) for dark energy, the cosmological constant $\Lambda$ is the most widely accepted model of dark energy due to the fact that the standard model of cosmology, so-called $\Lambda$CDM model, reconciles very well with the current observations. Instead of introducing dark energy, there is however an alternative to resolve this puzzle by modifying GR so that the dynamical behaviors of spacetime deviate from GR mainly at the cosmic scale, especially in such a way that the late-time universe coincides with that dominated by positive cosmological constant, i.e. de Sitter (dS) universe.

One of interesting modifications is to introduce a mass term for the graviton field.  Historically, adding the mass term in the Einstein's gravity can give theoretically undesirable consequences, including the ghost instability \cite{Boulware:1972yco}. Although there have been numerous attempts to formulate the models of ghost-free massive gravity, the most successful one is the de Rham, Gabadadze, and Tolley (dRGT) prescription in adding a combination of  mass terms in the Einstein-Hilbert action; these allowed mass terms, including quadratic, cubic and quartic ones, in the dRGT massive gravity give no higher derivative term in the equations of motion, resulting in the absence of ghost field \cite{deRham:2010kj,deRham:2010ik}. See \cite{deRham:2014zqa,Hinterbichler:2011tt} for review papers. Fortunately, the dRGT massive gravity can provide the solution whose parameters can be interpreted as a cosmological constant. As a viable model of gravity to address cosmological mysteries, there have been studies on the dRGT massive gravity in many respects.  These include several static and spherically symmetric black hole solutions in the dRGT massive gravity and their thermodynamic properties \cite{Capela:2011mh,Arraut:2014uza,Cai:2014znn,Ghosh:2015cva,Xu:2015rfa,Hu:2016mym,Zou:2016sab,EslamPanah:2016pgc,Hendi:2016hbe,Hendi:2016uni,Li:2016fbf,Hendi:2016yof,Hendi:2017fxp,Hendi:2017bys,Chabab:2019mlu}, accretion disk around a dRGT black hole \cite{Kazempour:2022asl}, greybody factor \cite{Boonserm:2017qcq,Kanzi:2020cyv,Boonserm:2021owk}, quasinormal modes \cite{Wongjun:2019ydo,Burikham:2020dfi}, black string solutions \cite{Tannukij:2017jtn,Ghosh:2019eoo} and their thermodynamics \cite{Sriling:2021lpr}, and constraining the model's parameters using the observational data \cite{Cardone:2012qq,Ponglertsakul:2018smo}, etc.

The dRGT massive gravity can have a black hole in the asymptotically background spacetime as anti-de Sitter (AdS) and dS, depending on the values of parameters. With a certain range of parameters that give rise to the dS-like universe, the black hole thermodynamics in the dRGT model is expected to provide several features as a natural extension of that of the dS black hole.    However, the discussion about the thermodynamic behaviors of a black hole in the asymptotically dS space is not tractable as desirable due to the nature of multi-horizon system. Usually, the Schwarzschild-de Sitter (Sch-dS) system has two horizons consisting of the black hole event horizon and the cosmological horizon.  Generically, the temperature at one horizon is not the same as one another, therefore the Sch-dS system is not in a thermodynamic equilibrium.  Due to its similarity in nature with the Sch-dS, the thermodynamic consideration of the dS black hole from the dRGT massive gravity also encounters the difficulties in applying the equilibrium thermodynamics due to its multi-horizon nature.

There have been some lessons from dealing with the Sch-dS black hole thermodynamics that can be used to apply in the dRGT black hole. The problem of multi-horizon system in the Sch-dS can be addressed by using either the separated system approach or effective system approach.  For the separated system approach, the system evaluated at each horizon can be defined independently \cite{Kubiznak:2015bya}.  For the effective system approach, the whole system can be considered as a single system in equilibrium \cite{Urano:2009xn}. The effective system approach can be done in two versions with considering the black hole mass $M$ as the internal energy and chemical enthalpy.  Considering the mass as the internal energy, the first law of thermodynamics can be in the form $dM = T_{eff}dS - P_{eff} dV$ \cite{Urano:2009xn,Ma:2013aqa,Zhao:2014raa,Ma:2014hna,Kubiznak:2016qmn}, where the total entropy and the volume can be defined as $S = S_{b} + S_{c}$ and $V = V_{c} - V_{b}$, respectively. Note that the subscripts $b$ and $c$ refer to the black hole event horizon and cosmological horizon. On the other hand, treating the mass as the enthalpy, the first law of thermodynamics of the effective system becomes $dM = T_{eff}dS + V_{eff} dP$, where the total entropy and the volume can be defined as $S = S_{b} + S_{c}$ and $V_{eff} = \left(\frac{\partial M}{\partial P}\right)_{S}$, respectively, with pressure $P \sim \Lambda$ \cite{Kubiznak:2016qmn,Li:2016zca}. Accordingly, the effective temperature of both versions can be defined as $T_{eff} = \left(\frac{T_{b}T_{c}}{T_{b} - T_{c}}\right)$. It is seen that the effective temperature blows up at the limit $T_{b} \rightarrow T_{c}$. This problem can be solved by using the new definition of the total entropy as $S = S_{b} - S_{c}$ \cite{Kanti:2017ubd,Chabab:2020xwr}.  Using this form of total entropy, the effective temperature can be defined as $T_{eff} = \left(\frac{T_{b}T_{c}}{T_{b} + T_{c}}\right)$, which does not blow up at the limit $T_{b} \rightarrow T_{c}$.  However, the entropy $S = S_{b} - S_{c}$ can be argued that it is not a physical entropy. In this work, we use the total entropy defined as $S = S_{b} + S_{c}$. The change of the total entropy can be investigated by considering that the direction of heat flow for the system evaluated at the cosmological horizon is opposite to one at the black hole horizon, since the observer stays between the black hole horizon and cosmological horizon \cite{Nakarachinda:2021jxd}. Consequently, the change of total entropy can be obtained as $dS = dS_{b} - dS_{c}$. From the expression of the change of total entropy, the effective temperature can be defined as $T_{eff} = \left(\frac{T_{b} T_{c}}{T_{b} + T_{c}}\right)$.
It is worth to apply these methods to explore the black hole thermodynamics in the dRGT massive gravity.

The black hole thermodynamics has been argued that it should be studied with non-extensive entropy as evident from the area law of the Bekenstein-Hawking entropy \cite{Bekenstein:1973ur,Gibbons:1977mu,Page:2004xp}. One of the generalized non-extensive entropy is proposed by Tsallis \cite{Tsallis:1987eu}.  Considering a system with two correlated subsystems, its Tsallis entropy satisfies the pseudoadditive composition rule
\begin{eqnarray} \label{Tsallis com}
S_{T}^{12}=S_T^1+ S_T^2 +\lambda S_T^1 S_T^2, 	
\end{eqnarray}
where $S_T^{12}$ is the Tsallis entropy of the entire system, $S_T^1$ and $S_T^2$ are the Tsallis entropies of the two separated subsystems, and $\lambda$ is the non-extensive parameter.
For one of the simplest choices, the black hole entropy can be thought of as the Tsallis entropy for the nonextensive system.  However, the definition of the empirical temperature using Tsallis entropy is not compatible with the zeroth law of thermodynamics \cite{Bir2011ZerothLC}.  To address this unclear definition of the temperature, using the formal logarithm, the Tsallis entropy can be transformed to the additive generalized entropy known as the R\'{e}nyi entropy \cite{Rnyi1959OnTD,Jizba:2002um}
\begin{eqnarray} \label{Renyi Log}
S_{R}^{12}=\frac{1}{\lambda} \ln \left[ 1+ \lambda S_T^{12}\right],  	
\end{eqnarray}
Thus, the empirical temperature cannot be defined as $T_{R} = \left(\frac{\partial M}{\partial S_{R}}\right)$ \cite{Bir2011ZerothLC}. Moreover, using the R\'enyi statistics instead of Gibbs-Boltzmann statistics,  the spherically symmetric black holes such as Sch black hole \cite{Czinner:2015eyk}, Sch-dS black hole \cite{Tannukij:2020njz}, rotating black hole \cite{Czinner:2017tjq} and charged black hole \cite{Promsiri:2022qin} have been found to be thermodynamically stable. In addition, the investigation of the black hole thermodynamics with R\'enyi entropy have been intensively considered  \cite{Biro:2013cra,Promsiri:2021hhv,Alonso-Serrano:2020hpb,Barzi:2022ygr,Cimdiker:2022ics,ElMoumni:2022chi}.

In this work, the thermodynamic stability of the black hole in asymptotically dS space in the dRGT massive gravity theory will be investigated using the R\'enyi statistics with both the separated system and effective system approaches. Since the dRGT black hole can provide corrections to the Sch-dS black hole, we analyze how the thermodynamic properties of the dRGT black hole are modified compared to the Sch-dS black hole. It is well known that either the thermodynamic pressure or volume of the Sch-dS black hole is negative. Actually, for positive thermodynamic volume, the ``pressure" $P= -\Lambda/(8\pi)$, the conjugate quantity to the volume, which is obviously negative for Sch-dS black hole should be viewed as tension rather than the pressure according to the standard thermal concepts. Therefore, this may be one of the difficulties in capturing the thermodynamic notion of the Sch-dS black hole. However, for the dRGT black hole, we find that it is possible to realize a black hole as a thermodynamic system whose pressure and volume can be chosen to be positive, thanks to the dRGT model parameters. In other words, dRGT massive gravity may provide a black hole which may be understood through the standard viewpoints of thermodynamics. An ability to realize positive pressure as well as positive volume is one of the worthy properties of the dRGT black hole compared to those in Sch-dS black holes.

With the separated system approach, the local stability of the black hole is analyzed by considering the sign of heat capacity.  Moreover, the lower bound of the parameter $\lambda$ for the local stability condition is determined. Furthermore, the global stability of the black hole can be analyzed by considering the Gibbs free energy. 
, which also yields a stronger lower bound on $\lambda$.
Finally, for the separated system, the phase transition between the non-black hole and black hole can be analyzed and the Hawking-Page phase transition is the first-order phase transition. For the effective system, the thermodynamic quantities can be defined by using the first law of thermodynamics as $dM = T_{eff}dS + V_{eff}dP$ where the mass $M$ is thought as the chemical enthalpy and the total entropy obeys the following addition rule, $S = S_{R_{1}} + S_{R_{2}}$, as seen in \cite{Sriling:2021lpr,Nakarachinda:2021jxd}. The local stability of the black hole can be analyzed by using the same steps as done in the separated system, from which the lower bound on $\lambda$ can be obtained. Furthermore, we find that there exists a particular range in temperature for which the black hole is locally stable either viewed through the effective system approach or separated system approach. As a result, if a black hole is observed to be at a temperature within this range, one may distinguish these two approaches by observing the size of the black hole. In the effective system, there is no lower bound on the nonextensive parameter determined through the global stability analysis. Eventually, the phase transition of the effective system between the non-black hole and the black hole can be analyzed. In particular, the Hawking-Page phase transition is the zeroth-order phase transition. This is one of the significant results which is different from the separated system.

This paper is organized as follows. In Sec.~\ref{sec:dRGT BH}, we review the dRGT black hole solution and then analyze its horizon structure. In Sec.~\ref{sec:thermo}, we investigate, on the former half of the section, thermodynamic properties of the black hole treated as two separated systems using R\'{e}nyi entropy while  the latter half is devoted to an investigation on the effective system and the thermodynamic properties according to R\'enyi statistics. Finally, in Sec.~\ref{sec:conclusion}, we conclude the investigation as well as give remarks on the effects of nonextensivity on the black hole in dRGT massive gravity.

\section{dRGT Black Hole}\label{sec:dRGT BH}
The massive gravity theories have been investigated 
 since 1939 Fierz and Pauli (FP) \cite{Fierz:1939ix}. As discussed in the previous section, there were many obstructions until 2010, the viable nonlinear massive gravity theory was proposed by  de Rham, Gabadadze, and Tolley \cite{deRham:2010kj,deRham:2010ik}. In this section, the dRGT massive gravity theory will be reviewed. The static spherically symmetric solution in dRGT massive gravity theory and the horizon structure of the dRGT black hole are discussed.

\subsection{dRGT Massive Gravity}
In this subsection, we review an important ingredient of dRGT massive gravity theory. This theory is free of the Boulware-Deser ghost by incorporating higher-order interaction terms into the Lagrangian. The dRGT Massive gravity action is the well-known Einstein-Hilbert action including suitable nonlinear interaction terms given by

\begin{equation}
S = \frac{1}{16\pi} \int d^{4}x \sqrt{-g} \Big[R + m_{g}^{2} \,\mathcal{U}(g,f)\Big],\label{dRGT action}
\end{equation}

where $R$ is the Ricci scalar. Note that we use the convention with $G=1$. The interaction terms include graviton mass, $m_{g}$, and the potential terms $\mathcal{U}$ expressed as 

\begin{equation}
\mathcal{U}=\mathcal{U}_{2} + \alpha_{3} \mathcal{U}_{3} + \alpha_{4} \mathcal{U}_{4},
\end{equation}
where

\begin{eqnarray}
\mathcal{U}_{2}&=&[\mathcal{K}]^{2} - [\mathcal{K}^{2}],\\
\mathcal{U}_{3}&=&[\mathcal{K}]^{3} - 3[\mathcal{K}][\mathcal{K}^{2}] + 2[\mathcal{K}^{3}],\\
\mathcal{U}_{4}&=&[\mathcal{K}]^{4} - 6[\mathcal{K}]^{2}[\mathcal{K}^{2}] + 8[\mathcal{K}][\mathcal{K}^{3}] + 3[\mathcal{K}^{2}]^{2} - 6[\mathcal{K}^{4}].
\end{eqnarray}

The parameters $\alpha_{3}$ and $\alpha_{4}$ are free parameters of the theory. The quantity $[\mathcal{K}^{n}]$ is the trace of the $n$-th power of the matrix

\begin{equation}
\mathcal{K}^{\mu}_{\hspace{0.2cm}\nu} = \delta^{\mu}_{\nu} - \big(\sqrt{g^{-1}f}\,\big)^{\mu}_{\hspace{0.2cm}\nu}.
\end{equation} 

$g_{\mu \nu}$ and $f_{\mu \nu}$ are the physical metric and the fiducial/reference metric, respectively. 
The fiducial metric contains the Stuckelberg scalar playing the role to restore the diffeomorphism invariance.  Note that the systematic construction of the potential terms provides the scalar mode of the theory acting similar to the scalar field in Galileon theory at the decoupling limit. Hence, the theory admits  5 degrees of freedom without the additional ghost mode.  

By varying the action \eqref{dRGT action} with respect to the physical metric $g^{\mu\nu}$, the dynamical field equations can be expressed as

\begin{equation}
G_{\mu \nu} + m_{g}^{2} \mathit{X}_{\mu \nu} = 0,
\label{Einstein equation}
\end{equation}
where $G_{\mu \nu}$ is the Einstein tensor and $\mathit{X}_{\mu \nu}$ is the effective energy-momentum tensor obtained from varying the potential $\mathcal{U}$. This effective energy-momentum tensor can be written in terms of the matrix $\mathcal{K}_{\mu \nu}$ as

\begin{equation}
\mathit{X}_{\mu \nu} = \mathcal{K}_{\mu \nu} - [\mathcal{K}]g_{\mu \nu} - \alpha\left(\mathcal{K}_{\mu \nu}^{2} - [\mathcal{K}]\mathcal{K}_{\mu \nu} + \frac{\mathcal{U}_{2}}{2}g_{\mu \nu}\right) + 3\beta \left(\mathcal{K}_{\mu \nu}^{3} - [\mathcal{K}]\mathcal{K}_{\mu \nu}^{2} + \frac{\mathcal{U}_{2}}{2}\mathcal{K}_{\mu \nu} - \frac{\mathcal{U}_{3}}{6}g_{\mu \nu}\right).
\end{equation}

The parameter $\alpha_{3}$ and $\alpha_{4}$ are redefined as

\begin{equation}
\alpha_{3} = \frac{\alpha - 1}{3},\quad \alpha_{4} = \frac{\beta}{4} + \frac{1 - \alpha}{12}.
\end{equation}
From the Bianchi identity of the Einstein tensor, $\nabla^\mu G_{\mu\nu}=0$, the effective energy-momentum tensor is also covariantly divergence-free

\begin{equation}
\nabla^{\mu} \mathit{X}_{\mu \nu} = 0,
\end{equation}
where $\nabla^{\mu}$ is the covariant derivative associated with the physical metric $g_{\mu\nu}$. These equations will be used in order to solve for the static and spherically symmetric solutions. The resulting solutions correspond to the black hole called dRGT black holes.

\subsection{dRGT black hole solution and horizon structure}\label{horizonanalysis}
In this section, we will review of the dRGT black hole solution. By considering the static and spherically symmetric spacetime, the metric contains four independent radial functions. Note that we cannot use the coordinate transformation to get rid of two functions since we have chosen the gauge choice via the Stueckelberg scalars. The solutions can be classified into two branches; the metric with off-diagonal components and the diagonal metric. In this consideration, we will focus on the diagonal solution. For this choice, there are only two independent radial functions. As a result, the general form of the metric tensor can be written as

\begin{equation}
ds^{2} = -n(r)dt^{2} + f^{-1}(r)dr^{2} + r^{2} d\Omega^{2},
\end{equation}
where $d\Omega^{2} = d\theta^{2} + \sin \theta^{2} d\phi^{2}$ is the line element on 2-sphere. It is important to note that the solution of the physical metric depends on the form of the fiducial metric. In principle, the choice does not affect the existence of the ghost, one can choose the form of the fiducial metric in order to obtain the proper solution of the physical metric. For example, from a cosmological
viewpoint, the physical metric does not admit a nontrivial flat cosmological solution with a Minkowski fiducial metric \cite{articlecos}, but it does for the open FLRW solution \cite{Gumrukcuoglu:2011ew}. Moreover, the first FLRW solution with arbitrary geometry exists when the FLRW fiducial metric is considered \cite{Gumrukcuoglu:2011zh}. By generalizing the form of the fiducial metric, nontrivial cosmological solutions can be obtained \cite{Chullaphan:2015ija}. In this consideration, let us choose the fiducial metric as \cite{Vegh:2013sk}

\begin{equation}
f_{\mu\nu} = \text{diag}\big(0,0,c^2,c^2\sin^2\theta\big),
\end{equation}

where $c$ is a constant. Substituting these ansatz to Eq. (\ref{Einstein equation}), one found that two functions can be related by a constant e.g. $n(r)=f(r)+C$. In order to reduce the solution to the usual form, the constant can be set as zero. As a result, the solution can be written as

\begin{eqnarray}
ds^{2} &=& -f(r)dt^{2} + f^{-1}(r)dr^{2} + r^{2} d\Omega^{2},\\
f(r) &=& 1 - \frac{2M}{r} - m_{g}^{2} (c_{2}r^{2} - c_{1}r - c_{0}),\label{horiz fn}
\end{eqnarray}
where $M$ is the Arnowitt-Deser-Misner mass of the black hole. $c_{0} = c^{2} (\alpha + 3\beta)$, $c_{1} = -c(1 + 2\alpha + 3\beta)$ and $c_{2} = -3(1 + \alpha + \beta)$.  In addition, the horizon function can be split into two branches as asymptotically de Sitter (dS) space for $m_{g}^{2}c_{2} > 0$ and asymptotically anti-de Sitter (AdS) space for $m_{g}^{2}c_{2} < 0$. Furthermore, it can be reduced to the Sch-dS/AdS black hole by setting $c_0=c_1=0$ and $m_g^2c_2=\Lambda/3$. 

It is important to note that there exists a nonlinear scale called the Vainshtein radius, $r_V \sim \big(\frac{M}{c_2m_g^2}\big)^{1/3}$, at which the solution reduces to the Schwarzchild (Sch) black hole for $r\ll r_V$ and corresponds to the dRGT black hole in asymptotically dS/AdS spacetime for $r\gg r_V$. This radius can be obtained by comparing the black hole mass term with the $c_2$ term. Moreover, it is found that there exists another nonlinear scale $r_1\sim\big(\frac{M}{c_1 m_g^2}\big)^{1/2}$ and $r_0 \sim \frac{M}{c_0 m_g^2}$ which is obtained by comparing the black hole mass term to the $c_1$ and $c_0$ terms, respectively. At this radius, the linear terms ($c_0$ and $c_1$ terms) become dominant contributions and give significant modifications.
In order to capture the significant contribution from each term, let us redefine the dimensionless parameters as follows

\begin{equation}
r = \left(\frac{M}{a_{2}}\right)x, \quad
c_{0} = \left(\frac{a_{0}}{m_{g}^{2}}\right), \quad
c_{1} = \left(\frac{a_{1}a_{2}^{2}}{Mm_{g}^{2}}\right), \quad
c_{2} = \left(\frac{a_{2}^{3}}{M^{2}m_{g}^{2}}\right).
\end{equation}

As a result, the horizon function in Eq. \eqref{horiz fn} can be rewritten as

\begin{equation}
f(x) = 1+a_{0}-a_{2}\left(\frac{2}{x}-a_{1}x+x^{2}\right).
\label{horizon function 2}
\end{equation}

It is important to note that the dimensionless variables $x$ is actually scaled by the Vainshtein radius $x= a_2 r /M = r/r_V$. In this context, the parameter $a_2 = M/r_V$  will characterize how the event horizon differs from the Vainshtein radius. Moreover, the parameters $a_1 = r^2_V/r^2_1$ and $a_0 = r_V/r_0$ will characterize the nonlinear scale comparing to $r_V$. In the limit $r_{1,0} \rightarrow \infty$, the parameter $a_1$ and $a_0$ will go to zero then the nonlinear scale is characterized by only Vainshtein radius. As a result,  the solution recovers the Sch-dS/AdS solution as setting $a_0=a_1=0$. In this study, we are interested only in the asymptotically dS black hole corresponding to $a_2>0$.

In order to analyze the horizon structure of the black hole, let us first consider the case of the Sch-dS black hole. The horizon function in Eq.~(\ref{horizon function 2}) for the Sch-dS black hole is simply expressed as 

\begin{equation}
f(x) = 1 - a_{2}\left(\frac{2}{x} + x^{2}\right).
\label{horizon function of dS}
\end{equation}

Since we are considering asymptotically dS spacetime, the horizon function $f(x)$ is a concave function and the maximum point can be evaluated from $\frac{df}{dx} = 0$. As a result, the value of $x$ at the extremum point of $f(x)$ is given by

\begin{equation}
x_{ex} = 1.
\label{extrema of dS}
\end{equation}

Substituting Eq. (\ref{extrema of dS}) to Eq. (\ref{horizon function of dS}), the extremum value of the horizon function is
\begin{equation}
f(x_{ex}) = 1 - 3a_{2}.
\end{equation} 
By requiring $f(x_{ex}) \geq 0$, the condition, in which the Sch-dS black hole has the horizon(s), is then written as
\begin{equation}
0 < a_{2} \leq 1/3.
\end{equation}

\begin{figure}[ht]\centering
\includegraphics[width=8cm,height=8cm]{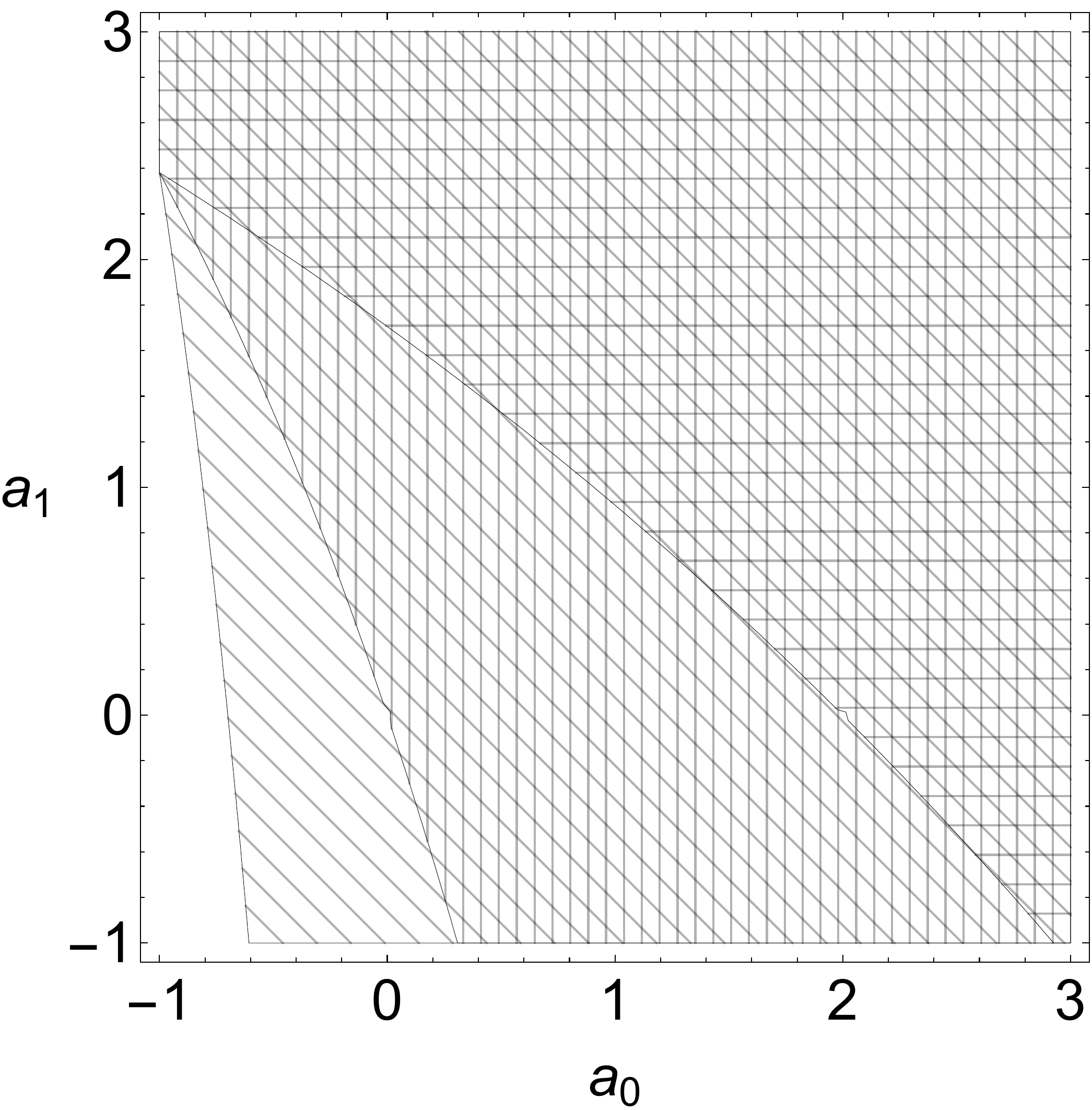}
\quad\quad
\includegraphics[width=8cm,height=8cm]{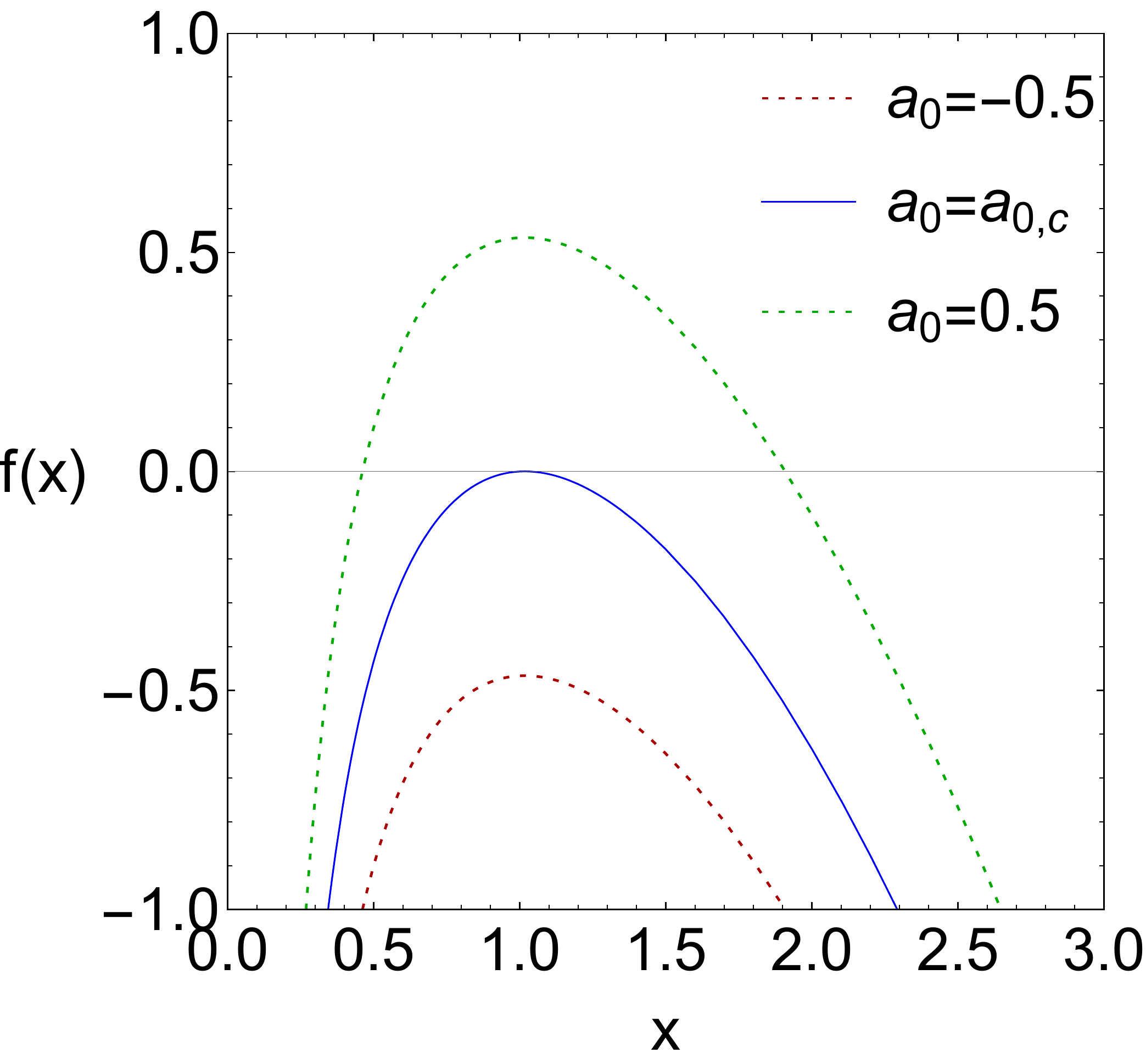}
\caption{Left panel shows the region of the existence of the horizons in $(a_{0},a_{1})$-space with specific value of $a_{2}$. The oblique, vertical and horizontal shading regions correspond to ones for $a_2=1/10$, $a_2=1/3$ and $a_2=1$, respectively. Right panel shows the behaviors of the horizon function $f(x)$ of the dRGT solution versus $x$ with various values of $a_{0}$ $(a_{0, c} = -0.0336)$ by fixing $a_{1}=1/10$, $a_{2} = 1/3$.}
\label{The horizon of dRGT}
\end{figure}

Now, let us consider the full expression of the dRGT solution. The maximum point can be obtained by using the same strategy as the one in the Sch-dS case. As a result, the value of $x$ at the maximum point of the horizon function is obtained as

\begin{equation}
x_{ex} = \frac{1}{6} \left[a_{1} + A_1 \left(1+  \frac{a^2_{1}}{A^2_1}\right) \right],
\label{extrema of horizon}
\end{equation}
where $A_{1} = 3\times 2^{2/3} \Big(1+ \frac{a_{1}^{3}}{108} - \sqrt{1 + \frac{a_{1}^{3}}{54}}\,\Big)^{1/3}$. Substituting Eq. (\ref{extrema of horizon}) to Eq. (\ref{horizon function 2}), the condition for having horizon can be obtained by using the requirement; $f(x_{ex}) \geq 0$. The maximum value $f(x_{ex})$ is lengthy, it is not convenient to show explicitly here. However, the condition for having the horizons can be illustrated by using a region plot as shown in the left panel of Fig. \ref{The horizon of dRGT}. From the right panel of this figure, it is seen that there exist the horizons even $a_{2} > 1/3$ with $a_{0}$ and $a_{1}$ are not zero. This is one of the important results compared to the Sch-dS solution. It is allowed to have horizons with the parameter range $a_2 > 1/3$.  For the case of a small value of $a_2$, one can perform the suitable approximation in order to properly find the deviation from the Sch-dS solution  since it is in the region for having two horizons as shown in the left panel of Fig. \ref{The horizon of dRGT} for the oblique shading region. 

Since the Sch-dS solution can have more than one horizon, this affects the thermal properties of the corresponding black hole. One of them is that the temperatures evaluated at each horizon are different from one another, which causes the black hole system to be out of the thermal equilibrium. In the next section, this problem will be treated through two different approaches: the separated system approach where each horizon is treated as two separated thermal systems, and the effective system approach where all the horizons are treated as a single effective thermal system.

\section{Thermodynamics}\label{sec:thermo}

From the previous section, with appropriate conditions on parameters $a_0,a_1,a_2$ as shown in Fig. \ref{The horizon of dRGT}, the dRGT black hole can form two event horizons. In order to explore thermal properties of the black hole, one may start with evaluating temperatures which, in general, are different for different horizons. This inevitably renders the black hole to be a non-equilibrium thermal system and the standard thermodynamics cannot be well applied. In order to do such investigation, one may consider the two horizons, with their individual temperatures and other thermal quantities, to be two separated thermal systems, each in quasi-equilibrium. On the other hand, one may collectively consider the two horizons as a single effective thermal system which is in thermal equilibrium. This section is dedicated to such two thermodynamical approaches.

\subsection{Separated system approach}\label{sepapp}

In this subsection, the thermodynamics of black hole in dRGT massive gravity is investigated by defining the thermodynamic quantities of each horizon separately. The mass $M$ can be found by solving $f(r_{h}) = 0$, where $r_{h}$ is the horizon of the black hole. As a result, the mass $M$ is obtained as

\begin{equation}
M = \frac{r_{h}}{2} \Big[1 - m_{g}^{2}\big(c_{2} r_{h}^{2} - c_{1}r_{h} - c_{0}\big)\Big].
\label{parameter mass 1}
\end{equation}

The Hawking temperature of the dRGT black hole can be obtained from the surface gravity, $\kappa$, evaluated at the horizon as follows:

\begin{equation}
T_{b,c} \equiv \frac{\kappa_{b,c}}{2\pi}=\frac{|f'(r_{b,c})|}{4\pi} = \pm \frac{\big[1 - m_{g}^{2}(3c_{2}r_{b,c}^{2} - 2c_{1}r_{b,c} - c_{0})\big]}{4 \pi r_{b,c}},
\label{Hawking temperature}
\end{equation}
where the subscripts $b$ denote quantities evaluated at the black hole horizon, like $r_{b}$, and the subscripts $c$ denote those evaluated at the cosmological horizon, like $r_{c}$. Here, the plus and minus signs in Eq. (\ref{Hawking temperature}) denote the temperature of the system evaluated at $r_{b}$ and $r_{c}$, respectively.

\begin{figure}[ht]\centering
\includegraphics[width=8cm,height=8cm]{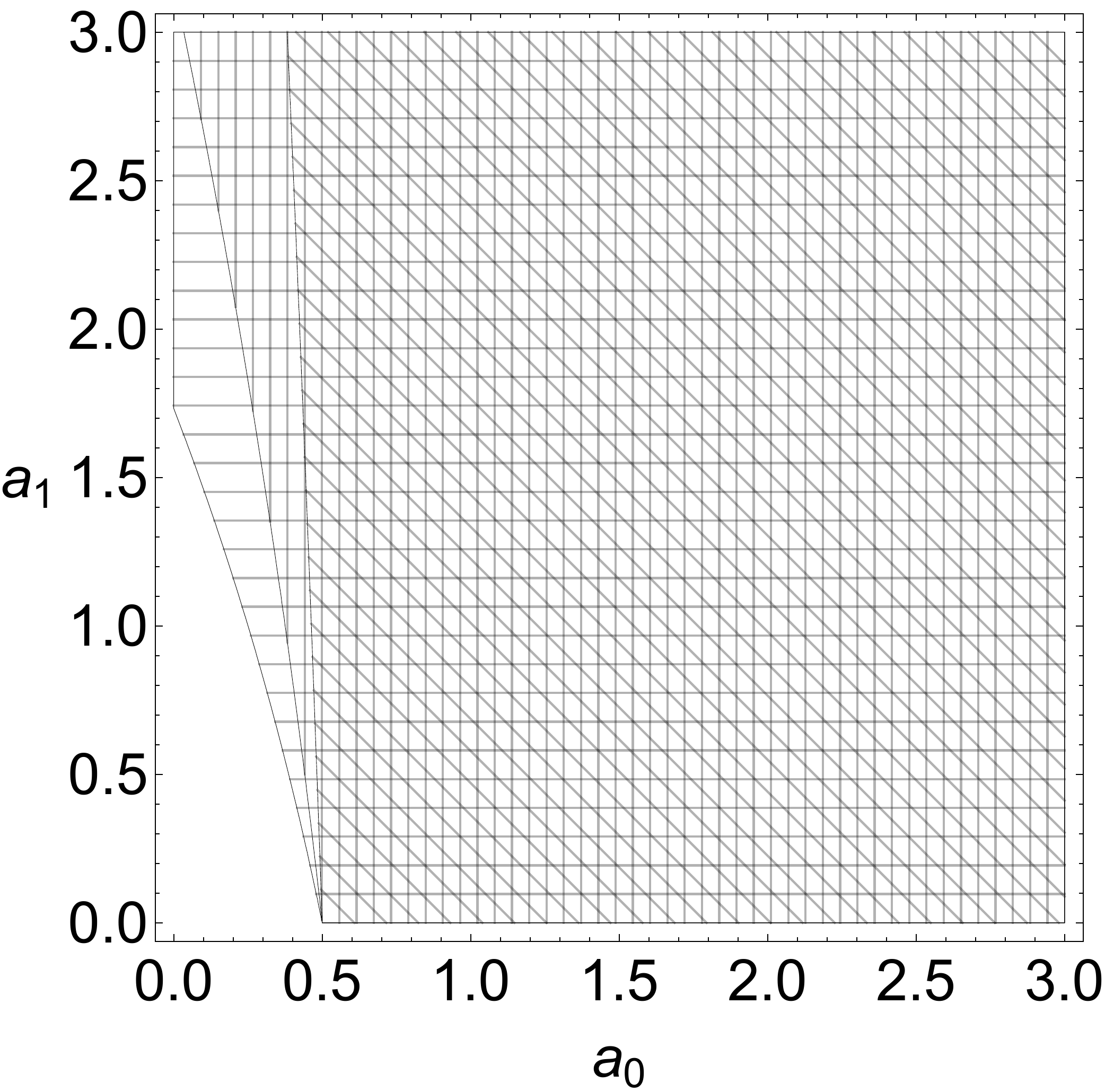}
\caption{The figure shows the region of the existence of the positive volume in $(a_{0},a_{1})$-space with various values of $a_{2}$. The oblique, vertical and horizontal shading regions correspond to one for $a_2 = 1/3$, $a_2 = 1/10$ and $a_2 = 1/100$, respectively.}
\label{The volume of dRGT}
\end{figure}

The entropy of the system corresponding to the temperature defined in Eq. \eqref{Hawking temperature} is given by using the Bekenstein-Hawking entropy, $S_{BH}$ as

\begin{equation}
S_{BH} = \frac{A}{4},
\end{equation}
where $A =4 \pi r_{h}^{2}$ is the surface area of the horizon of the black hole. The mass, $M$, temperature, $T_{b,c}$, and Bekenstein-Hawking entropy, $S_{BH}$, satisfy first law of thermodynamics as $dM = \pm T_{b,c}dS_{BH}$. The first law of thermodynamics can be extended by treating the other parameters as thermodynamic variables. In order to generalize the first law of thermodynamics, let us consider the Smarr formula of the black hole by treating the mass $M$ from Eq. (\ref{parameter mass 1}) as the homogeneous function. The mass $M$ is said to be a homogeneous function of thermodynamic quantities, $S,m_{g}^{-2},c_{0},c_{1}^{2}$, if it satisfies the following relation.

\begin{equation}
M(JS,Jm_{g}^{-2},Jc_{0},Jc_{1}^{2}) = J^{1/2} M(S,m_{g}^{-2},c_{0},c_{1}^{2}),
\label{parameter mass 2}
\end{equation}
where $J \in R$ and the function $M$ is said to be homogeneous of order $1/2$. By using the Euler's theorem, the Smarr formula can be written by using Eq. (\ref{parameter mass 2}) as

\begin{equation}
M = \pm 2S_{BH} T_{b,c} - 2PV_{b,c} + 2c_{0} \Phi_{0} + c_{1} \Phi_{1},
\label{Smarr formula}
\end{equation}
where $P \equiv \frac{3}{8 \pi} m_{g}^{2}$. The conjugates to the thermodynamic quantities can be identified via the Euler's theorem  as follows:

\begin{eqnarray}
T &=& \pm \left(\frac{\partial {M}}{\partial S}\right)_{r_{b,c}} = T_{b,c}, 
\label{Hawking temperature 2}\\
V_{b,c} &=& \left(\frac{\partial {M}}{\partial P}\right)_{r_{b,c}} = \frac{4}{3} \pi r^{3}_{b,c} \left(\frac{c_{0}}{r^{2}_{b,c}} + \frac{c_{1}}{r_{b,c}} - c_{2}\right), \label{Vbc}\\
\Phi_{0} &=& \left(\frac{\partial {M}}{\partial c_{0}}\right)_{r_{b,c}} = \frac{4}{3} \pi P r_{b,c}, \\
\Phi_{1} &=& \left(\frac{\partial {M}}{\partial c_{1}}\right)_{r_{b,c}} = \frac{4}{3} \pi P r^{2}_{b,c}.
\end{eqnarray}
Note that the temperature in Eq. (\ref{Hawking temperature 2}) is the same as in Eq. (\ref{Hawking temperature}). Furthermore, it is possible for the black hole in dRGT massive gravity to have positive thermodynamic volume as well as positive thermodynamic pressure, if an appropriate set of parameters is assumed for Eq. \eqref{Vbc}. In particular, there exists a viable range of parameters corresponding to the positive thermodynamic volume as shown in the Fig. \ref{The volume of dRGT}.

However, for the black hole in GR with the cosmological constant, either volume or pressure will be negative as follows:

\begin{equation}
\displaystyle
\begin{array}{lr}
\displaystyle P = \pm \frac{\Lambda}{8\pi},& \displaystyle V = \mp \left(\frac{\partial M}{\partial P}\right).
\end{array}
\end{equation}
Eventually, the first law of thermodynamics of the black hole in dRGT massive gravity can be written as

\begin{equation}
dM = \pm T_{b,c}dS_{BH} + V_{b,c}dP + \Phi_{0}dc_{0} + \Phi_{1}dc_{1}.
\label{The first law 1}
\end{equation}
If one consider that $P$, $c_{0}$, and $c_{1}$ are fixed, then the first law of thermodynamics can be reduced to $dM = \pm T_{b,c}dS_{BH}$. In our work, we consider the case where the parameters $c_{0}$ and $c_{1}$ are fixed. Thus, the first law of thermodynamics can be written as

\begin{equation}
dM = \pm T_{b,c}dS_{BH} + V_{b,c}dP.
\label{The first law 2}
\end{equation}
Note that the first law in Eq. \eqref{The first law 2} is a result from assuming that the entropy of the black hole is that of Bekenstein-Hawking entropy which is proportional to the surface area of the black hole itself. This means the hole's entropy is not an extensive quantity. As mentioned in Sec. \ref{intro}, in order for one to study the black hole as an extensive thermal object, one may instead use thermodynamics based on the R\'{e}nyi statistics. To this end, $S_{BH}$ is treated to obey the Tsallis composition rule. In order to realize such system as an extensive thermal object, the formal logarithm of $S_{BH}$, the so-called R\'{e}nyi entropy, is considered as an entropy representing the system. Thus, the thermodynamics of the black hole can be studied by using the R\'{e}nyi entropy as

\begin{equation}
S_{R} = \frac{1}{\lambda} \ln(1 + \lambda S_{BH}),
\end{equation}
where $\lambda$ is the non-extensive parameter and $-\infty < \lambda < 1$. In order to restrict the R\'{e}nyi entropy so that it is always positive, it is sufficient to choose $0 < \lambda < 1$. Note that, the R\'{e}nyi entropy reduces to the Bekenstein-Hawking entropy when $\lambda \rightarrow 0$. The first law of thermodynamics based on R\'{e}nyi statistics is assumed to be

\begin{equation}
dM = \pm T_{R(b,c)}dS_{R(b,c)} + V_{b,c}dP,\label{1st law sep app}
\end{equation} 
where $T_{R(b)}$ and $T_{R(c)}$ represent the R\'{e}nyi temperatures corresponding to the system evaluated at the black hole horizon and the cosmological horizon, respectively, and  the thermodynamic pressure is defined as $P = \frac{3}{8\pi} m_{g}^{2}$. Applying the R\'{e}nyi entropy instead of $S_{BH}$, the R\'{e}nyi temperature can be obtained as follows:

\begin{equation}
T_{R(b,c)} = \pm \left(\frac{\partial {M}}{\partial {S_{R(b,c)}}}\right)_{P} = (1 + \pi \lambda r_{b,c}^{2})T_{b,c}.
\end{equation}
Let us define a dimensionless temperature in terms of dimensionless variables as follows

\begin{eqnarray}
\overline{T}_{R(b)} &=& r_V T_{R(b)} = \frac{(1 + a_{0} + 2a_{1}a_{2}x - 3a_{2}x^{2})(x^{2} + \epsilon)}{4\pi \epsilon x},\\
\overline{T}_{R(c)} &=& r_V T_{R(c)} = -\frac{(1 + a_{0} + 2a_{1}a_{2}y - 3a_{2}y^{2})(y^{2} + \epsilon)}{4 \pi \epsilon y},
\end{eqnarray}
where $r_{b} = r_V x$, $r_{c} = r_V y$, and $\lambda = \frac{1}{\epsilon \pi r_V^2}$. 
Note that the valid values of the black hole horizon and the cosmological horizon radii are in the ranges $0 < r_{b} \leq r_{c}$ and $r_{b} \leq r_{c} < \infty$, respectively. With the dimensionless variables, the mentioned ranges can be written as $0 < x \leq \frac{1}{3}a_{1} \left(1 + \sqrt{1 + \frac{3(1 + a_{0})}{a_{1}^{2} a_{2}}}\,\right)$ and $\frac{1}{3}a_{1} \left(1 + \sqrt{1 + \frac{3(1 + a_{0})}{a_{1}^{2} a_{2}}}\,\right) \leq y < \frac{1}{2}a_{1} \left(1 + \sqrt{1 + \frac{4(1 + a_{0})}{a_{1}^{2} a_{2}}}\,\right)$. Additionally, the black hole becomes extremal when $x = y = \frac{1}{3}a_{1} \left(1 + \sqrt{1 + \frac{3(1 + a_{0})}{a_{1}^{2} a_{2}}}\,\right)$.

\begin{figure}[ht]\centering
\includegraphics[width=8cm,height=8cm]{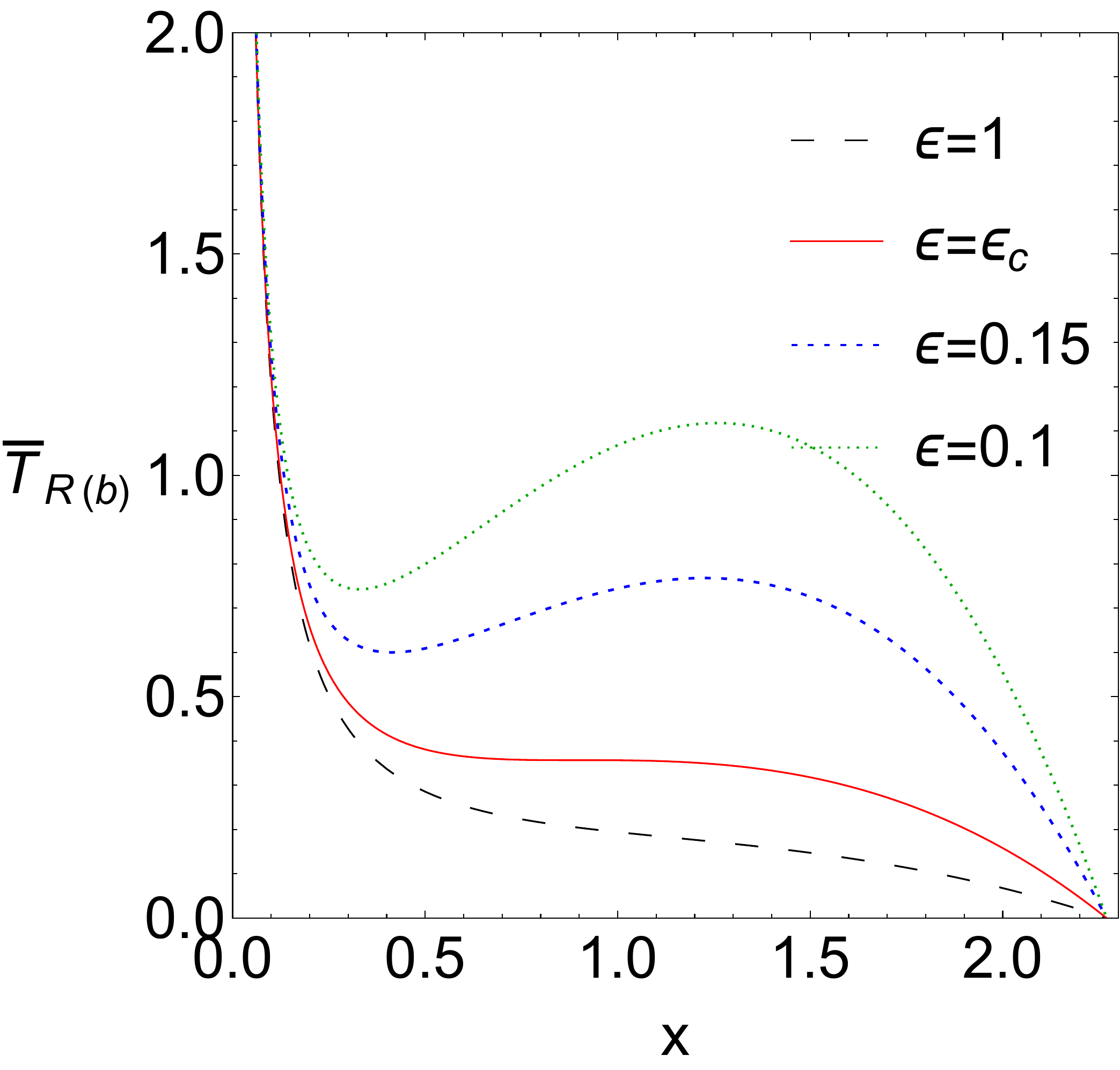}
\quad
\includegraphics[width=8cm,height=8cm]{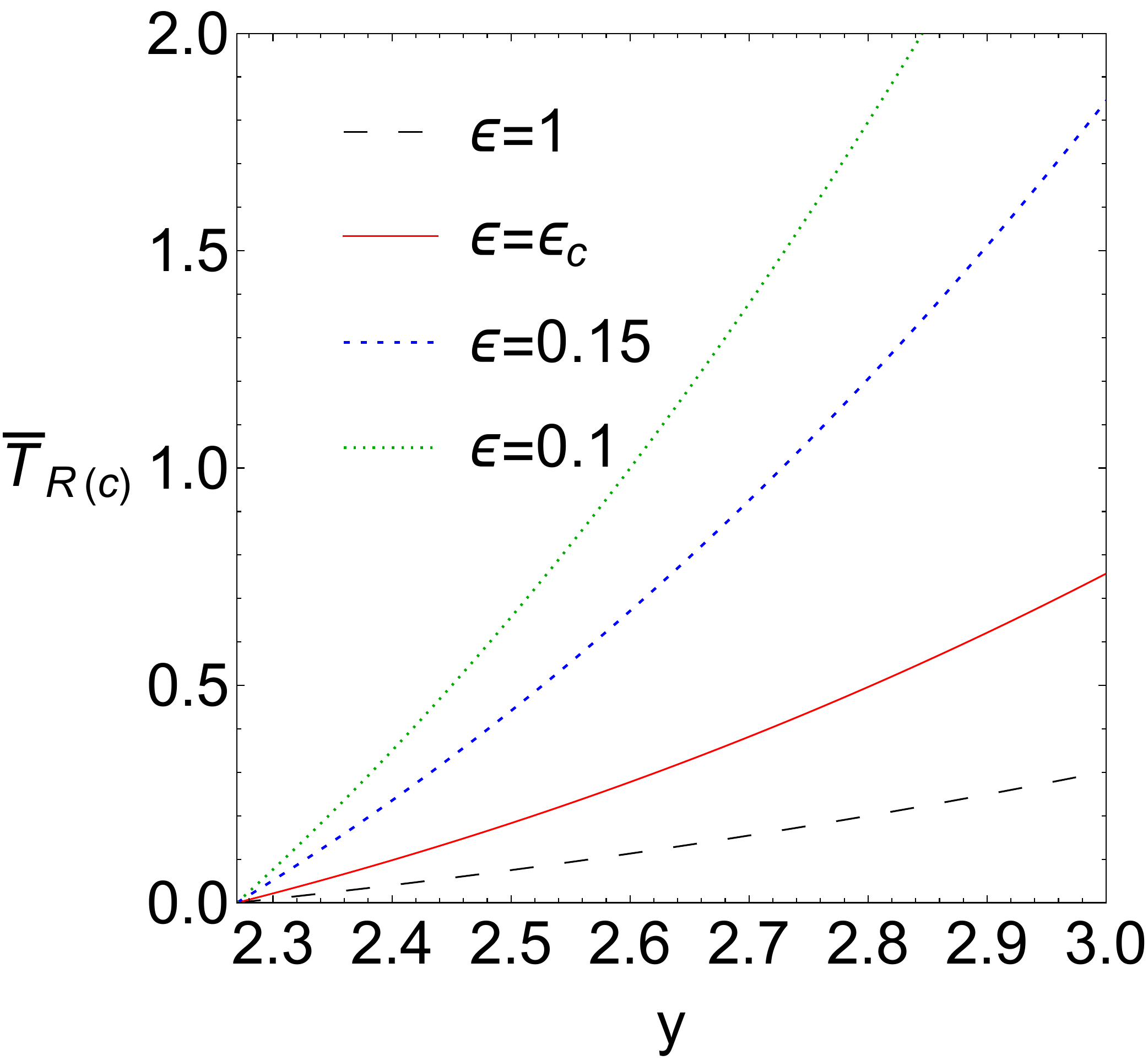}
\caption{Left (Right) panel shows the temperature of the system evaluated at the black hole (cosmological) horizon with various values of $\epsilon$ $(\epsilon_{C} = 0.37507)$ by fixing $a_{0} = 1/2$ and $a_{1} = 0.1 = a_{2}$.}
\label{temperature of separated}
\end{figure}

The R\'{e}nyi temperatures for each horizon, $\overline{T}_{R(b)}$ and $\overline{T}_{R(c)}$,  are shown explicitly in Fig. \ref{temperature of separated}. From this figure, one can see that there exists a range for positive slope implying the positive heat capacity. We will see later that the sign of heat capacity will directly relate to the slope of the temperature. As a result, in order to find the condition to obtain the positive positive heat capacity, we analyze the slope of the temperature. From Fig. \ref{temperature of separated}, $\overline{T}_{R(b)}$ exhibits two extrema while the profile of $\overline{T}_{R(c)}$ does not. Both extrema of $\overline{T}_{R(b)}$ can be found through its derivative as

\begin{equation}
F_{b} \equiv \frac{d \overline{T}_{R(b)}}{dx} = \frac{-9a_{2}x^{4} + 4a_{1}a_{2}x^{3} + (1 + a_{0} - 3a_{2}\epsilon)x^{2} - (1 + a_{0})\epsilon}{4 \pi \epsilon x^{2}}.
\label{extrema of temperature b}
\end{equation} 
For the positive value of $x$, the graph of $F_{b}$ is  concave. There are two real roots for Eq. (\ref{extrema of temperature b}). The extremum point, $x_{b}$ of the function $F_{b}$, i.e. the turning point of $\overline{T}_{R(b)}$, can be obtained by solving $\frac{d F_{b}}{dx}=0$. Then, by substituting $x_{b}$ in the function $F_{b}$, the locally stable condition on the non-extensive parameter can be found by requiring that the slope at the turning point of $\overline{T}_{R(b)}$ vanishes, or $F_{b}(x_{b})=0$. As a result, one can obtain the local bound on the nonextensive parameter as $\epsilon_{C} = \epsilon_{C}(a_{0},a_{1},a_{2})$. Note that the subscript $C$ denotes the bound corresponding to the heat capacity (being positive). The expression for $\epsilon_{C}(a_{0},a_{1},a_{2})$  is too lengthy and not necessary to be expressed explicitly here. However, it may be useful to approximate $\epsilon_C$ in order to study its features. In the case of $a_{1}$ and $a_{2}$ being negligibly small, one obtains the local stability condition on the nonextensive parameter as

\begin{figure}[ht]\centering
\includegraphics[width=8cm,height=8cm]{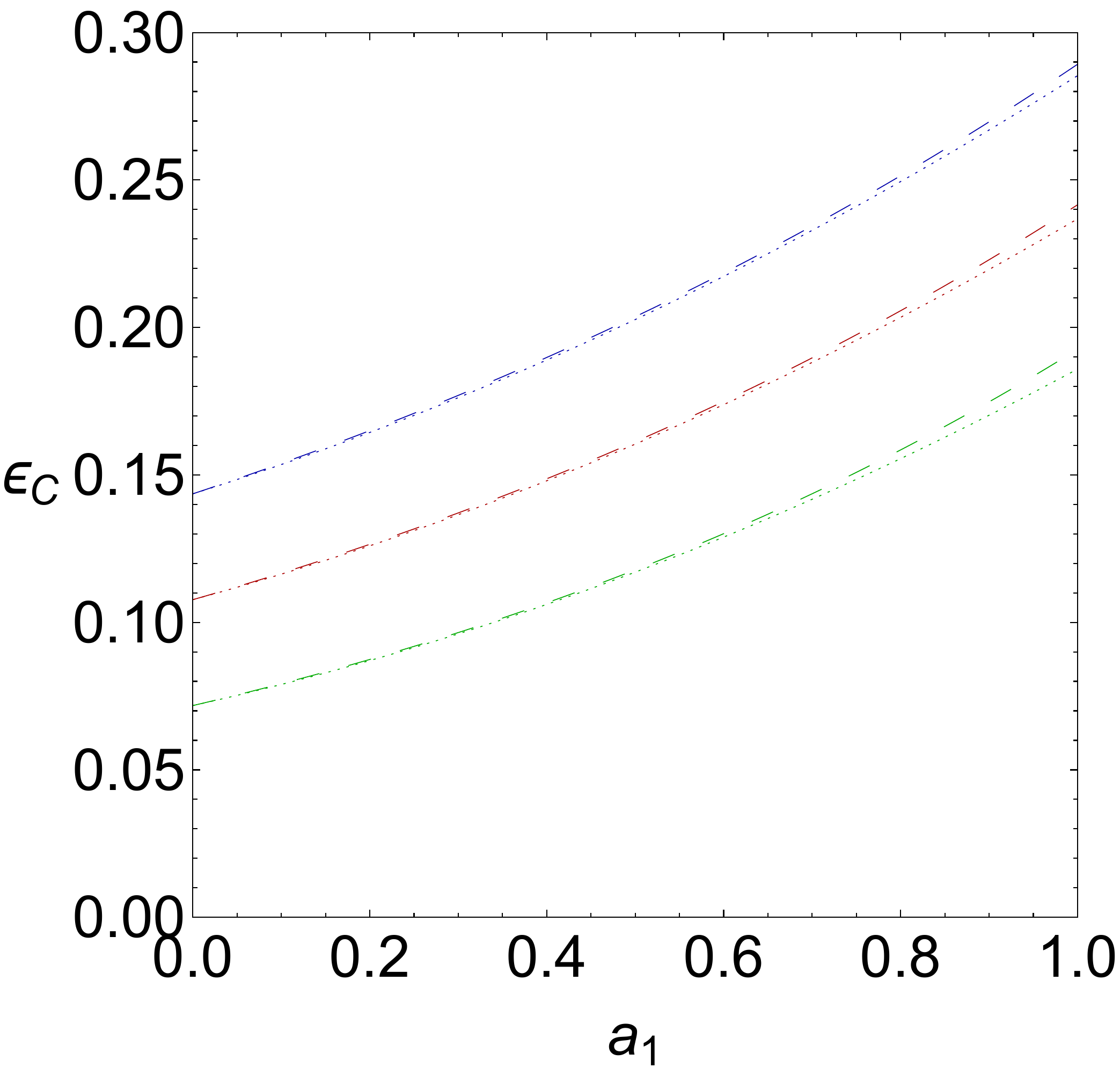}
\quad
\includegraphics[width=8cm,height=8cm]{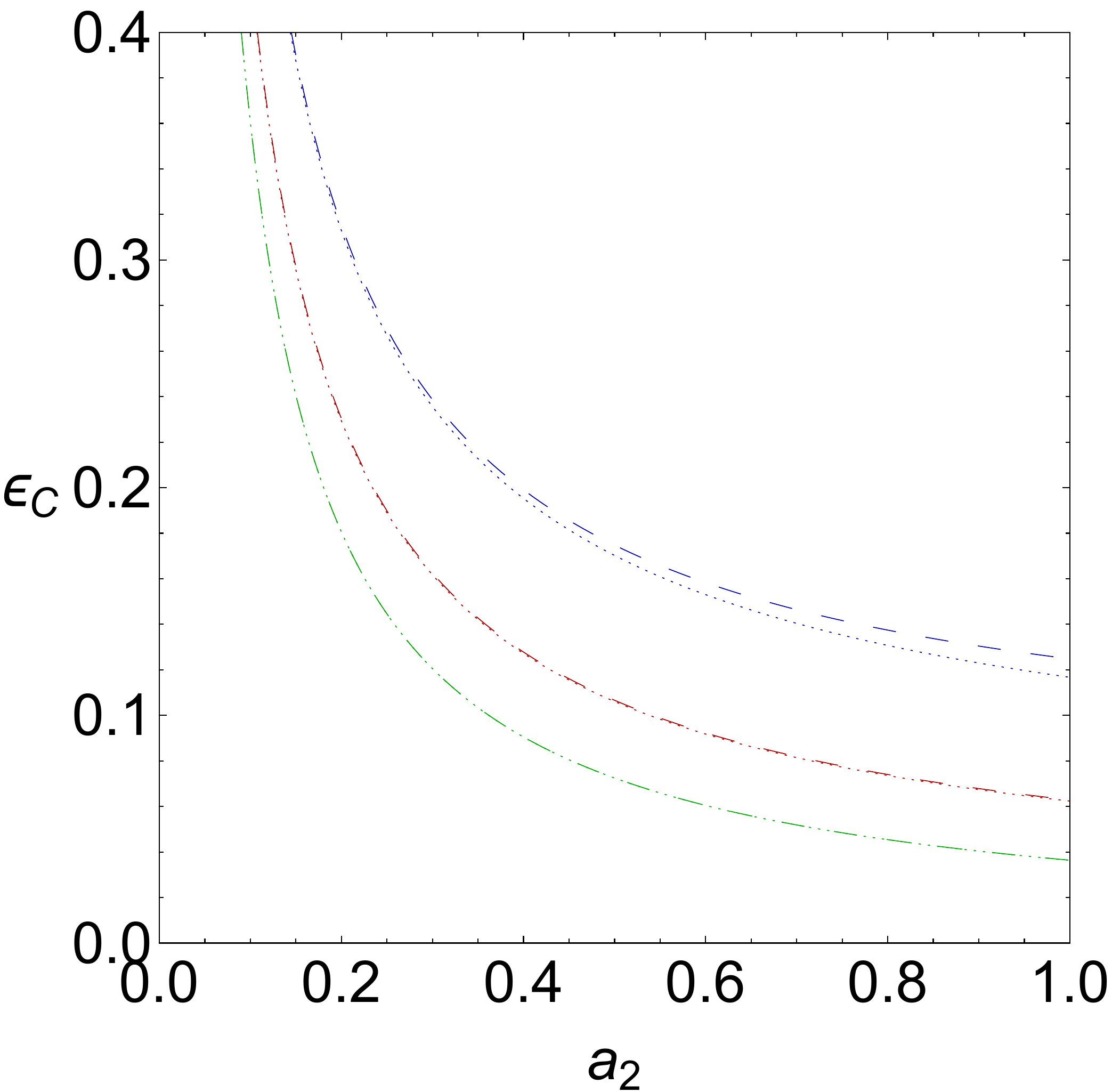}
\caption{Left panel shows the comparison of $\epsilon$ for the local stability in the full version (dashed line) and approximation (dotted line) versus $a_{1}$ by fixing $a_{2} = 1/4$ (blue), $a_{2} = 1/3$ (red) and $a_{2} = 1/2$ (green) with $a_{0} = 1/2$. Right panel shows the comparison of $\epsilon$ for the local stability in the full version (dashed line) and approximation (dotted line) versus $a_{2}$ by fixing $a_{1} = 0.9$ (blue), $a_{1} = 0.4$ (red) and $a_{1} = 0.01$ (green) with $a_{0} = 1/2$.}
\label{The local stability of dRGT}
\end{figure}

\begin{equation}
\epsilon \leq \epsilon_{C} \approx  \frac{\epsilon_{C(dS)} }{3 a_2}\left[1 + \sqrt{\frac{13}{5}} \frac{(a_{1} \sqrt{a_{2}} + a_{1}^{2} a_{2})}{\sqrt{1 + a_{0}}}\right] (1 + a_{0}),
\label{condition of local}
\end{equation} 
where $\epsilon_C$ and $\epsilon_{C(dS)}$ denote the upper bounds for the dRGT and Sch-dS black holes, respectively. Note that the value of $\epsilon_{C(dS)}$ is expressed numerically as
$\epsilon_{C(dS)}=7 - 4\sqrt{3}\approx 0.0718$ \cite{Tannukij:2020njz}. 
From Fig. \ref{The local stability of dRGT}, it can be seen that the exact upper bound, $\epsilon_C$, is more than the approximated upper bound. This suggests that although Eq. \eqref{condition of local} is an approximated expression, it serves well as a borderline to the condition on the existence of locally stable systems. Note that, by setting $a_{0} = 0 = a_{1}$, we obtain $\epsilon_{C(dS)} = 3 a_2 \epsilon_C$. The factor $3a_2$ appears due to the fact that we rescale the radial coordinate by $r_V$ instead of $L_{mg} \sim 1/m_{g}$ while the number $3$ will be gotten rid of by setting $m^2_g = \Lambda/3$. In this limit, there exists the nonextensivity length $L_\lambda \sim 1/\sqrt{\lambda}$ which may relate to the fine-graining parameter as argued in \cite{Nakarachinda:2021jxd, Promsiri:2021hhv},
\begin{equation}
\displaystyle
\frac{L_\lambda}{L_{mg}} \leq  \sqrt{7 - 4\sqrt{3}} \approx 0.268.
\end{equation}
This equation shows that nonextensivity length must be small enough compared to $L_{mg}$ to obtain the locally stable black hole. From a cosmological viewpoint, the length scale of graviton mass is proportional to the Hubble radius $L_{mg} \sim H_0^{-1}$. This means in order to stabilize small black holes, compared to the Hubble radius, the nonextensivity should be taken into account.

For the dRGT black hole case, there are correction terms corresponding to nonlinear effects at radius $r_1$ and $r_0$. For setting $a_0 \neq 0$ and $a_1 = 0$, one can see that the correction term is proportional to the parameter $a_0$ characterized by the nonlinear scale $r_0 = r_V/a_0 $. For $a_0 < 1$, we have $r_0 > r_V$ implying that between $r_V$ and the Hubble radius, there exists a length scale $r_0$ which modifies the bound of the nonextensivity due to the structure of graviton mass. In addition to length scale $r_0$, there is nonlinear scale $r_1$ which can be obtain by setting $a_1 \neq 0$ and $a_0 = 0$. From this setting, the leading contribution can be expressed as $a_1 \sqrt{a_2}$. In order to capture some physical meaning of this contribution, let us consider the horizon $r_{h}$ scaled as $r_{h} \sim a_1 r_V$ and nonextensive length scaled $L_\lambda \sim r_V/\sqrt{a_2}$. As a result, one obtains

\begin{equation}
\displaystyle
\frac{r_{h}}{L_\lambda} \sim a_1 \sqrt{a_2} .
\end{equation}
One can see that if the black hole horizon is comparable to the nonextensive length $r_h \sim a_1 \sqrt{a_2}$, the correction terms become dominant while if it is small, we can neglect these corrections.

For the thermodynamic system at the cosmological horizon, the slope of temperature is always positive for $\frac{1}{3}a_{1} \left(1 + \sqrt{1 + \frac{3(1 + a_{0})}{a_{1}^{2} a_{2}}}\,\right) < y < \frac{1}{2}a_{1} \left(1 + \sqrt{1 + \frac{4(1 + a_{0})}{a_{1}^{2} a_{2}}}\,\right)$. It is also possible to find extrema of $\overline{T}_{R(c)}$ by solving $\frac{d \overline{T}_{R(c)}}{dy} = 0$. However, the extrema are out of the valid range of $y$. Therefore, there are no extrema for the temperature of the system evaluated at $r_{c}$. The behavior of the temperature at $r_{c}$ can be shown in the right panel of Fig. \ref{temperature of separated}.

\begin{figure}[ht]\centering
\includegraphics[width=8cm,height=8cm]{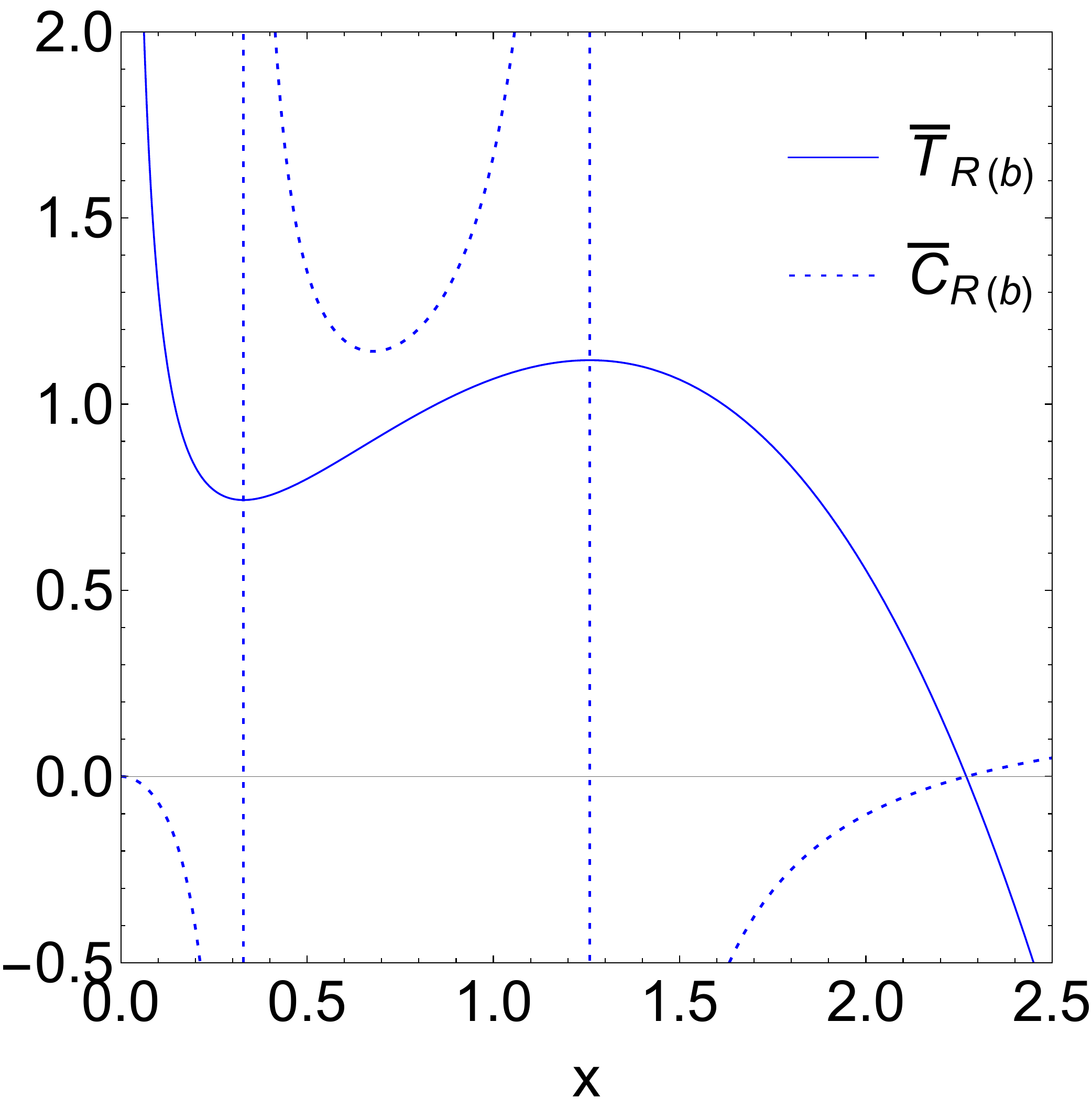}
\quad
\includegraphics[width=8cm,height=8cm]{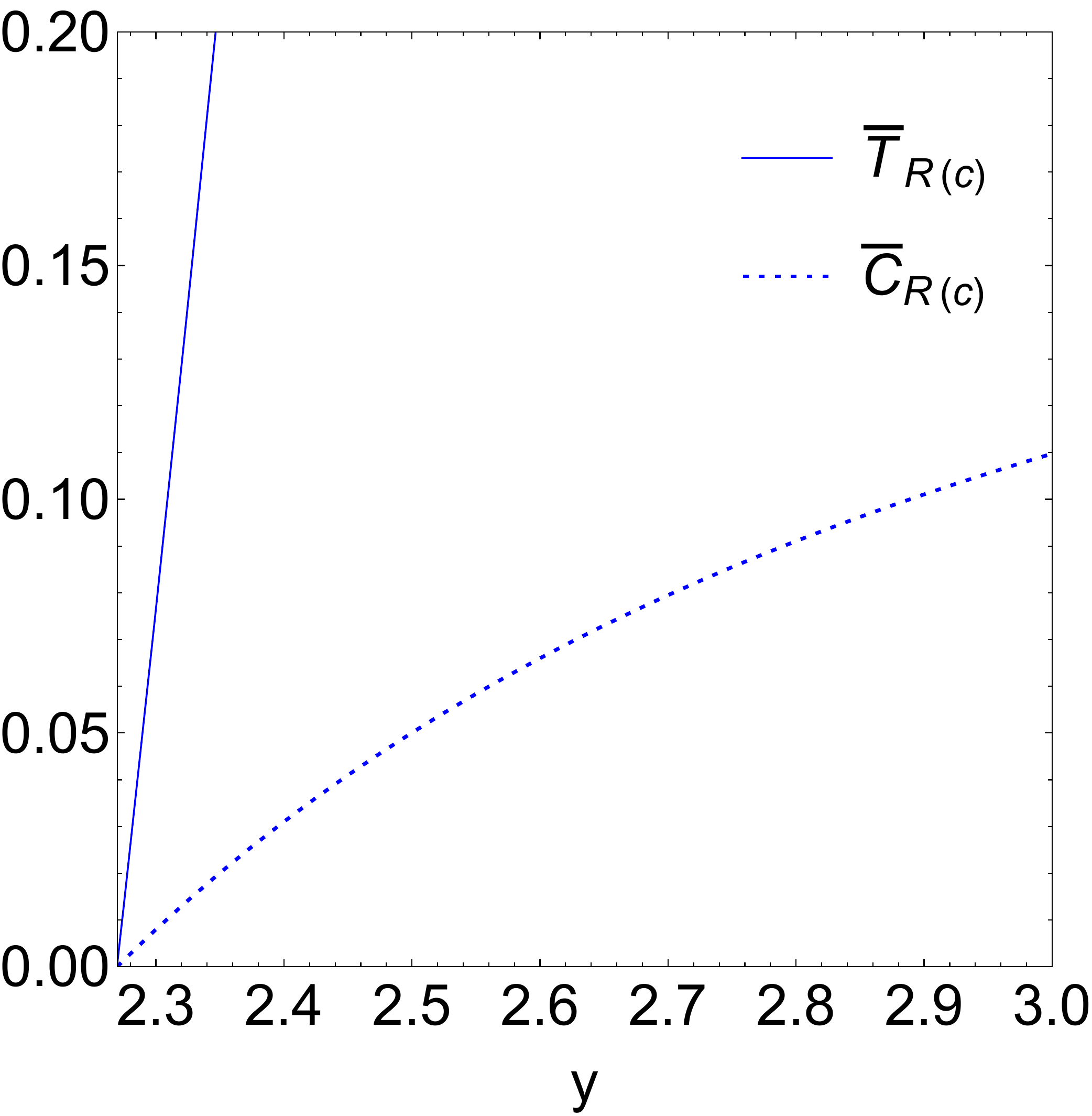}
\caption{Left (Right) panel shows the heat capacity and temperature of the system evaluated at the black hole (cosmological) horizon with the parameters are set as $a_{0} = 1/2$, $a_{1} = 0.1 = a_{2}$ and $\epsilon=0.1$.}
\label{heat capacity of separated}
\end{figure}

The black hole system can also be considered in terms of its local thermal stability. In particular, the system is said to be locally stable if its heat capacity is positive. Otherwise, it will radiate thermal radiation. Eventually, the black hole will vanish. In other words, the black hole with negative heat capacity is locally unstable. The heat capacity with fixing $P$ is defined as

\begin{eqnarray}
&&C_{R(b,c)} = \pm \left(\frac{\partial {M}}{\partial {T_{R(b,c)}}}\right)_{P}, \label{CfixP}
\\
\overline{C}_{R(b)} = \frac{C_{R(b)}}{r_V^2}  &=& \frac{2\pi x^{2} (-1-a_{0}-2a_{1}a_{2}x+3a_{2}x^{2})\epsilon}{9a_{2}x^{4} - 4a_{1}a_{2}x^{3} - (1+a_{0}-3a_{2}\epsilon)x^{2} + (1+a_{0})\epsilon},
\label{heat b}
\\
\overline{C}_{R(c)} = \frac{C_{R(c)}}{r_V^2} &=& \frac{2\pi y^{2} (-1-a_{0}-2a_{1}a_{2}y+3a_{2}y^{2}) \epsilon}{9a_{2}y^{4} - 4a_{1}a_{2}y^{3} - (1+a_{0}-3a_{2}\epsilon)y^{2} + (1+a_{0})\epsilon}.
\label{heat c}
\end{eqnarray}
These heat capacities are written in terms of dimensionless parameters. For the heat capacity of the system evaluated at black hole horizon, it can be shown explicitly in the left panel of Fig. \ref{heat capacity of separated}. The denominator of Eq. (\ref{heat b}) is the same as one in Eq. (\ref{extrema of temperature b}). Hence, the heat capacity diverge at the extrema of the temperature of the system, namely, $x_{-}$ and $x_{+}$. Moreover, this means the heat capacity is inversely proportional to the slope of the temperature, $C_{R(b)}\propto \frac{1}{\left(\frac{dT_{R(b)}}{dr_b}\right)}$, as can be seen explicitly in the left panel of Fig. \ref{heat capacity of separated}. There are three ranges of $x$ for the heat capacity of the system evaluated at black hole horizon: the smaller black hole whose size is smaller than the local minimum, $x_{-}$, the larger black hole whose size is bigger than the local maximum, $x_{+}$, and the moderate-sized black hole whose size lies within the range $x_{-}<x<x_{+}$. According to their heat capacities, the moderate-sized black hole is locally stable while the smaller and larger black holes are locally unstable. 

For the heat capacity of the system evaluated at cosmological horizon, there are no divergent points for $\overline{C}_{R(c)}$ in the ranges of $\frac{1}{3}a_{1} \left(1 + \sqrt{1 + \frac{3(1 + a_{0})}{a_{1}^{2} a_{2}}}\,\right) < y < \frac{1}{2}a_{1} \left(1 + \sqrt{1 + \frac{4(1 + a_{0})}{a_{1}^{2} a_{2}}}\,\right)$. The behavior of the heat capacity of the system evaluated at the cosmological horizon can be shown in the right panel of Fig. \ref{heat capacity of separated}. It is obviously seen that the heat capacity of the system evaluated at the cosmological horizon is always positive, which implies local stability of the system.

Apart from the local stability, one may also consider the global stability of this black hole system. The global stability can be analyzed by the Gibbs free energy as follows

\begin{figure}[ht]\centering
\includegraphics[width=8cm,height=8cm]{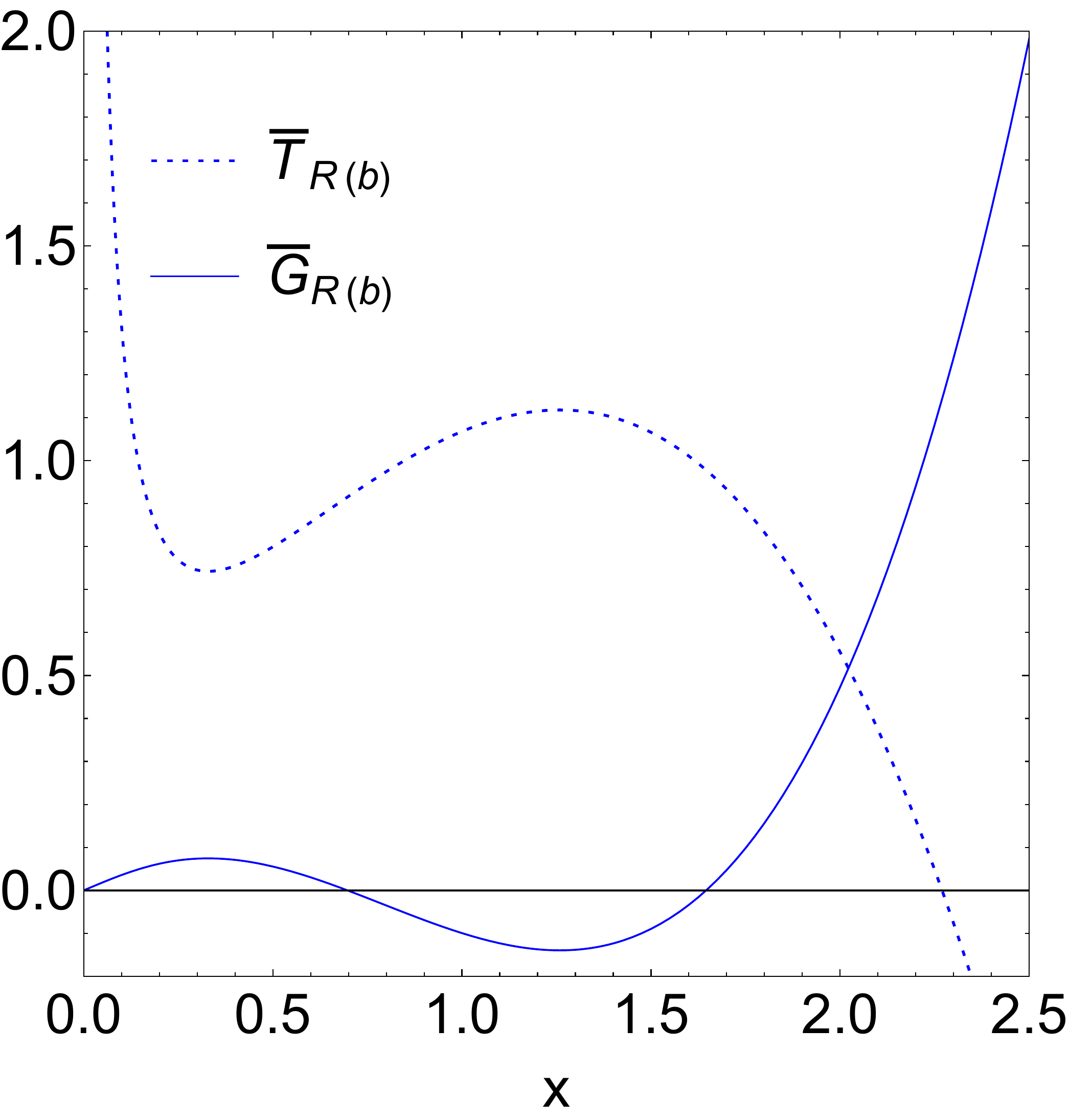}
\quad
\includegraphics[width=8cm,height=8cm]{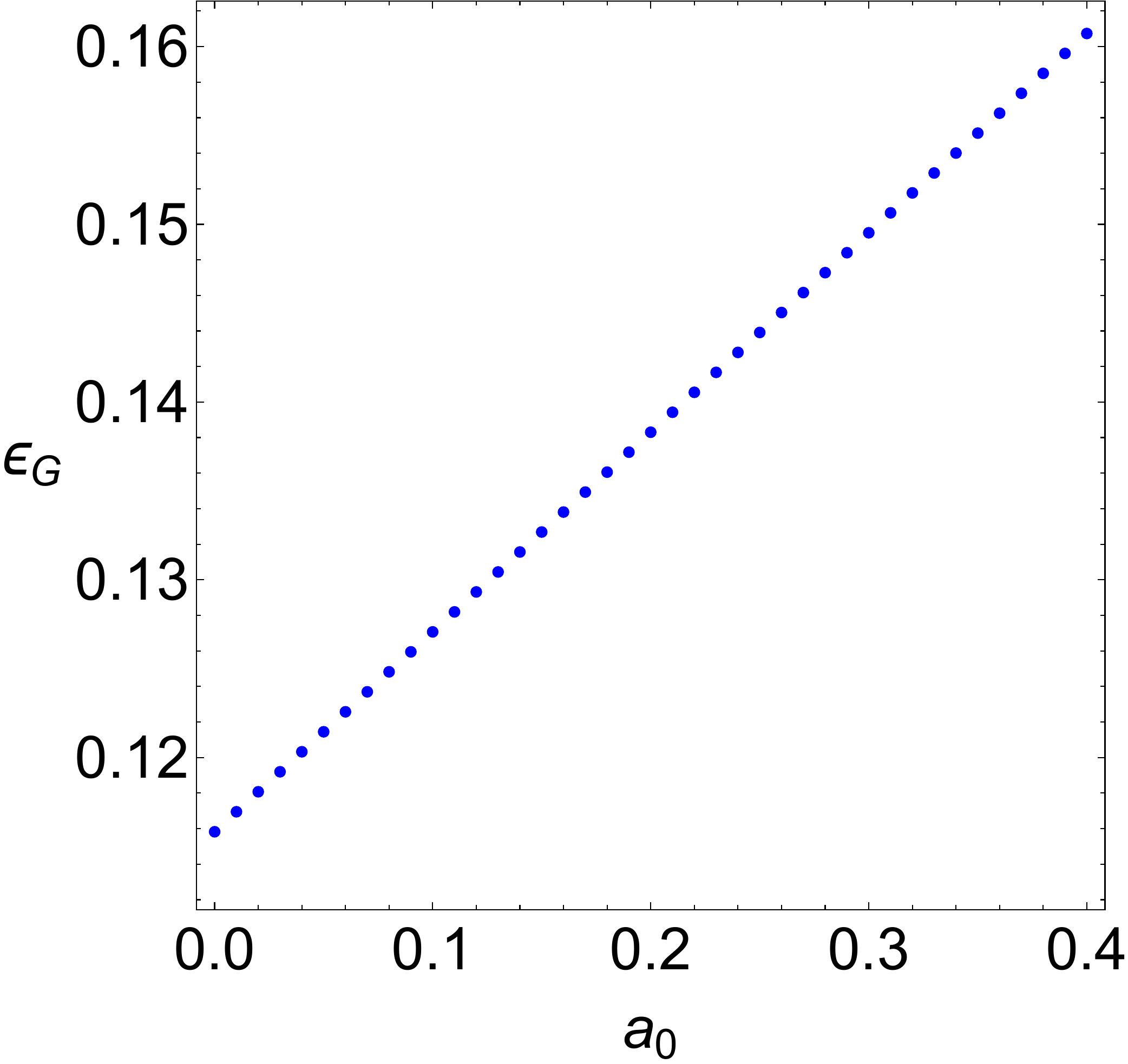}
\caption{Left panel shows the Gibbs free energy and temperature versus $x$ for the parameter setting $a_{0} = 1/2$, $a_{1} = 0.1 = a_{2}$ and $\epsilon = 0.1$. Right panel shows the behavior of $\epsilon_{G}$ versus $a_{0}$ by fixing $a_{1} = 0.1 = a_{2}$.}
\label{The plot of Gibbs and temperature}
\end{figure}

\begin{eqnarray}
&&\overline{G}_{R(b,c)}
=\frac{G_{R(b,c)}}{r_V}
=\frac{1}{r_V}\Big(M-T_{R(b,c)}S_{R(b,c)}\Big),
\\
\overline{G}_{R(b)}
&=& \frac{x}{2}\big[1+a_{0}+a_{2}x(a_{1}-x)\big]-\big[1+a_{0}+a_{2}x(2a_{1}-3x)\big]\left(\frac{x^{2}+\epsilon}{4x}\right) \ln \left(\frac{x^{2}+\epsilon}{\epsilon}\right),
\label{Gibbs b}
\\
\overline{G}_{R(c)}
&=& \frac{y}{2}\big[1+a_{0}+a_{2}y(a_{1}-y)\big]+\big[1+a_{0}+a_{2}y(2a_{1}-3y)\big]\left(\frac{y^{2}+\epsilon}{4y}\right) \ln \left(\frac{y^{2}+\epsilon}{\epsilon}\right).
\label{Gibbs c}
\end{eqnarray}
The system with global stability prefers the negative Gibbs free energy. The local minimum/maximum of the Gibbs free energy is the same points as the maximum/minimum of the temperature as seen on the left-hand side in Fig. \ref{The plot of Gibbs and temperature}. Therefore, the upper bound for $\epsilon$ can be found by requiring the condition $\overline{G}_{R(b)}(x_{+}) = 0$. In principle, the condition on $\epsilon$ can be written in terms of  $a_{0}$, $a_{1}$ and $a_{2}$. However, it is not easy to solve for the analytic solution. This problem is due to the logarithmic function in the Gibbs free energy. To obtain the global bound on the nonextensive parameter denoted by $\epsilon_G$, we may use the numerical method to show that the behavior of $\epsilon_{G}$ is linear function of the parameter $a_{0}$ as shown on the right-hand side in Fig. \ref{The plot of Gibbs and temperature}. Therefore  $\epsilon_{G}$ can be expressed in terms of $a_0, a_1$ and $a_2$ as

\begin{figure}[ht]\centering
\includegraphics[width=8cm,height=8cm]{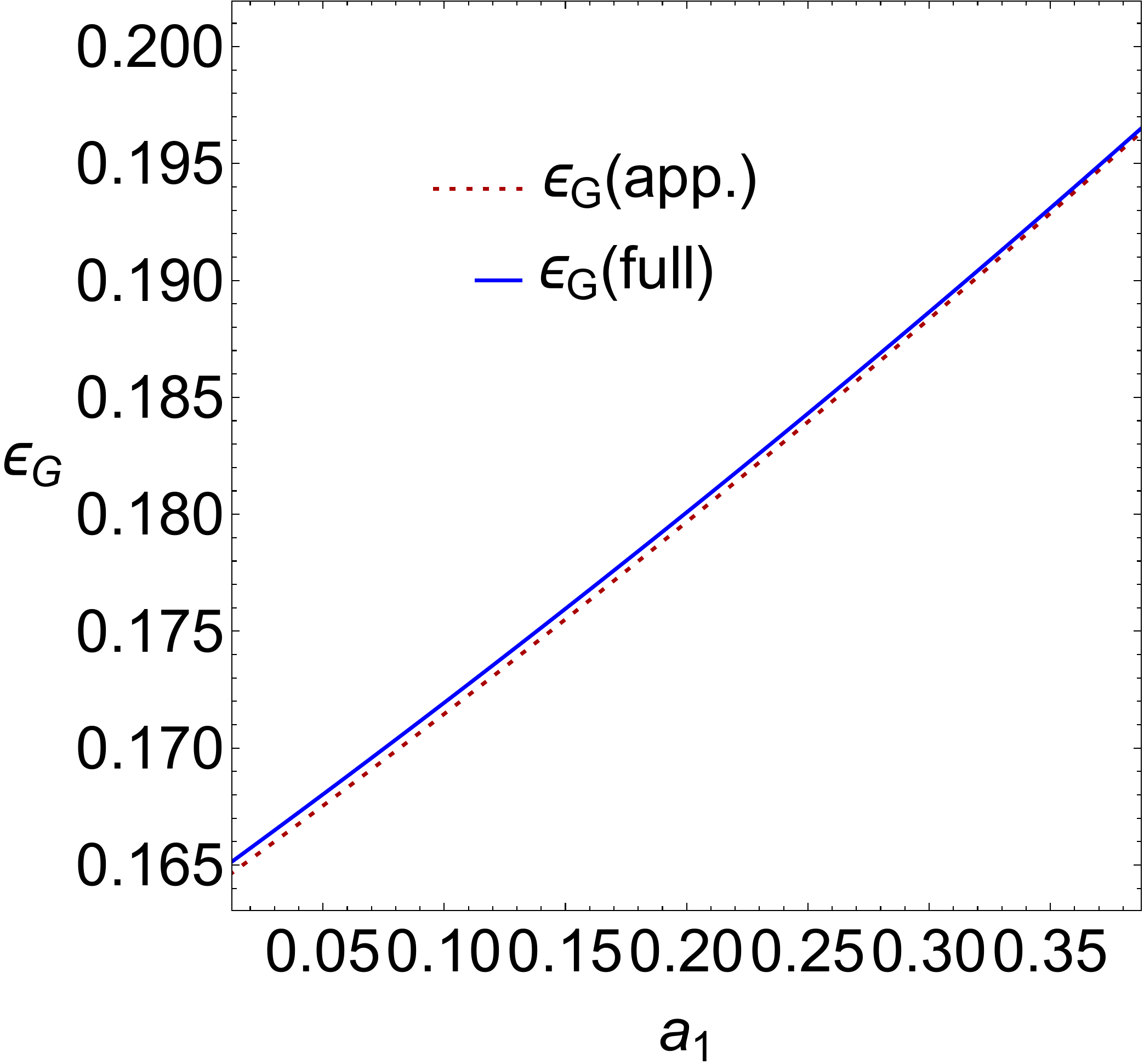}
\quad
\includegraphics[width=8cm,height=8cm]{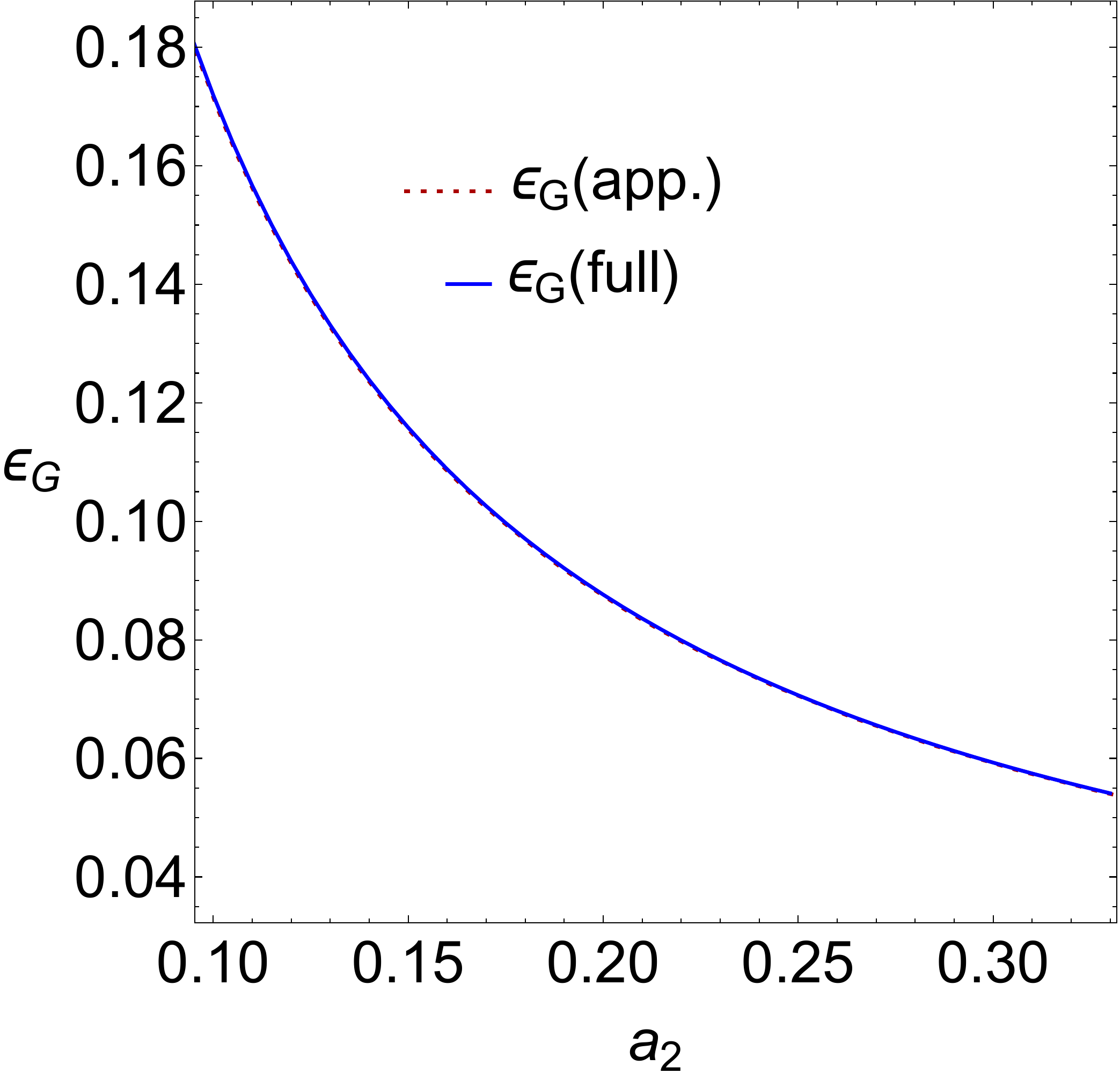}
\caption{Left (Right) panel shows the comparison of $\epsilon_{G}$ obtained from $\overline{G}_{R(b)}$ in full equation and approximation versus $a_{1}$ ($a_2$) by fixing $a_{0}=1/2$ and $a_{2} = 0.1$ ($a_1=0.1$).}
\label{comparison a1 and a2}
\end{figure}
\begin{figure}[ht]\centering
\includegraphics[width=7.8cm,height=7.2cm]{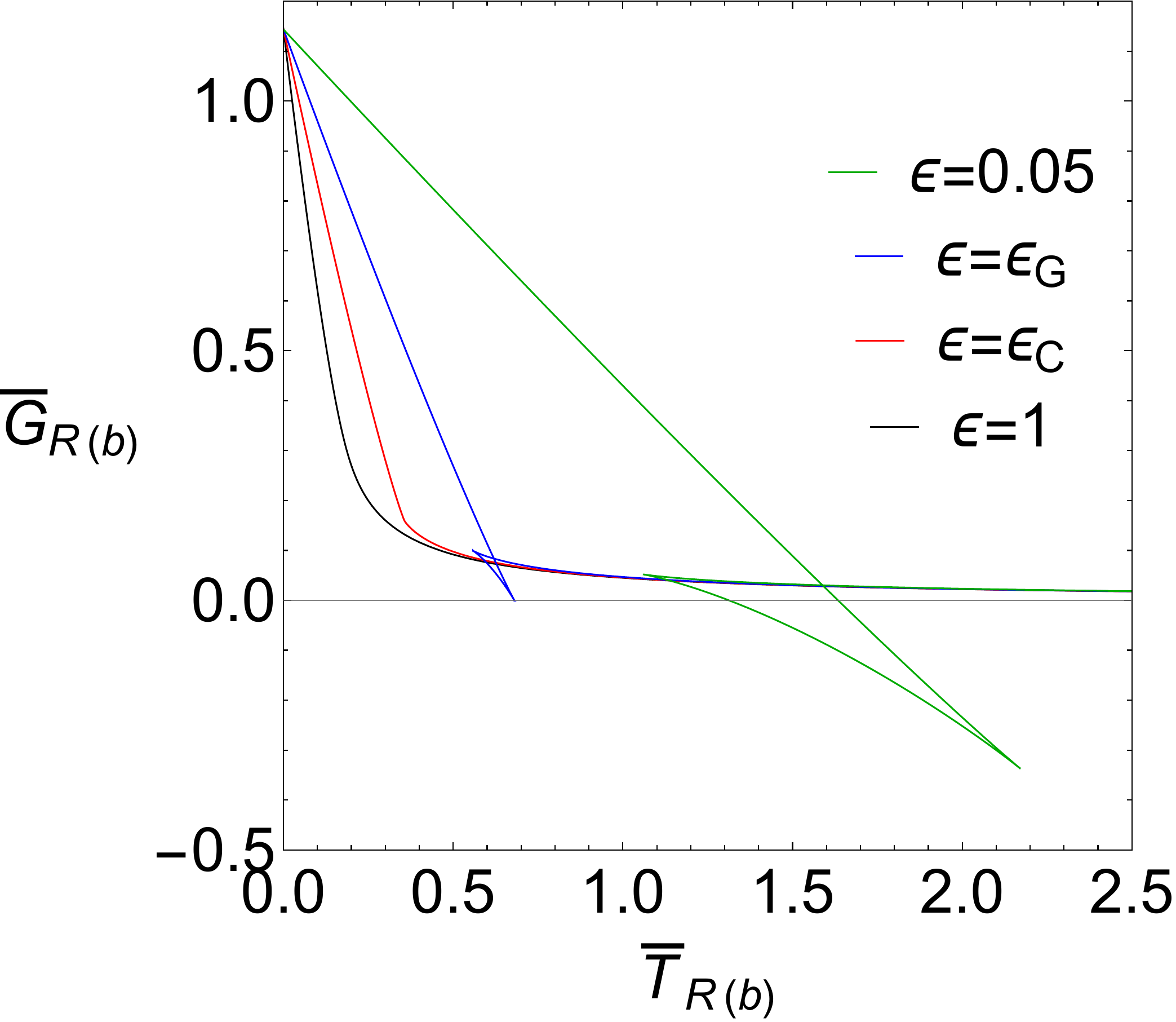}
\quad
\includegraphics[width=7.8cm,height=7.2cm]{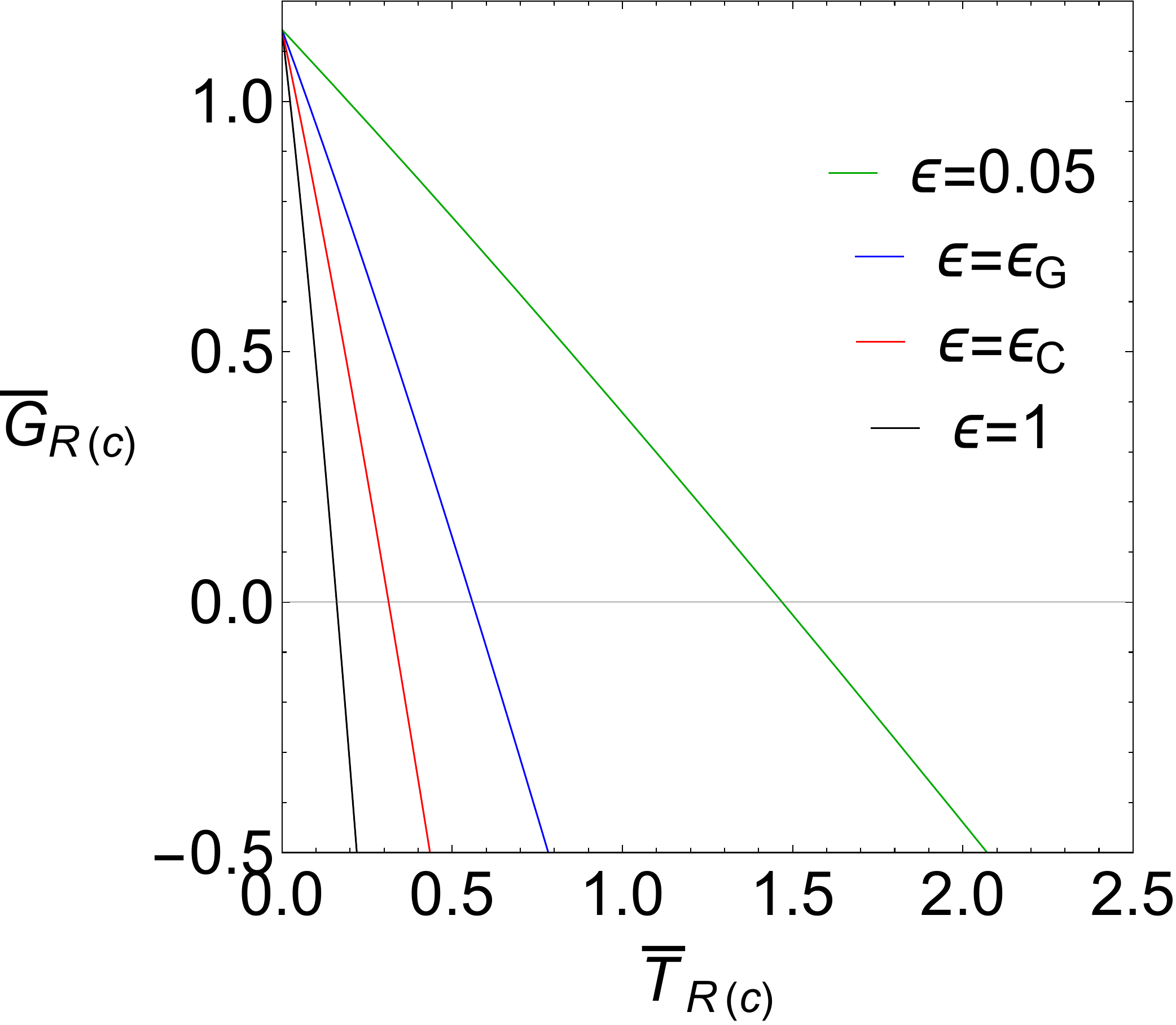}
\caption{Left (Right) panel shows the Gibbs free energy versus temperature of the system evaluated at the black hole (cosmological) horizon with various values of $\epsilon$ by fixing $a_{0} = 1/2$, $a_{1} = 0.1 = a_{2}$. Note that $\epsilon_{C} = 0.37507$, and $\epsilon_{G} = 0.17082$.}
\label{effective Gibbs}
\end{figure}


\begin{equation}
\epsilon_{G} \equiv f(a_{1},a_{2}) + g(a_{1},a_{2}) a_{0}.
\end{equation}
In order to find $f(a_{1},a_{2})$ and $g(a_{1},a_{2})$ for $\epsilon_G$ from above expression, we perform numerical method evaluating point by point. As a result, the condition on $\epsilon$ for the black hole to have the global stability while keeping $a_{1}$ and $a_{2}$ small is obtained as

\begin{eqnarray}
\epsilon \leq \epsilon_{G} &\approx& \frac{\epsilon_{G(dS)}}{3 a_2} \left[\left\{1 + \frac{1}{5} \sqrt{79} \left(a_{1} \sqrt{a_{2}} + a_{1}^{2} a_{2}\right)\right\} + \left\{1 + \frac{1}{5} \sqrt{\frac{78}{5}} \left(a_{1} \sqrt{a_{2}} + a_{1}^{2} a_{2}\right)\right\} a_{0}\right],\,\,
\label{condition of global}\\
\epsilon_{G(dS)}&=&  \sqrt{ \frac{26}{125}} \epsilon_{C(dS)}=\sqrt{ \frac{26}{125}} (7 - 4\sqrt{3})  \approx 0.0328,
\end{eqnarray}
where $\epsilon_{G(dS)}$ is the upper bound on $\epsilon$ due to the global stability analysis in the Sch-dS black hole \cite{Tannukij:2020njz}. 
From  Fig. \ref{comparison a1 and a2}, one can see that the approximated expression of $\epsilon_{G}$ in Eq. \eqref{condition of global}  is closed to the exact value of $\epsilon_{G}$ obtained from $\overline{G}_{R(b)}=0$. The behavior of the Gibbs free energy can be analyzed by using the relation $S_R = -\left(\frac{\partial G_{R}}{\partial T_{R}}\right)$. This implies that the slope of the graph $G_{R}-T_{R}$ is always positive. The behavior of Gibbs free energy with  different values of $\epsilon$ is illustrated  in the left panel of Fig. \ref{effective Gibbs}. From this figure, there exist two cusps corresponding to two extremum points in the temperature profile denoted by $x_{\pm}$. At these points, the heat capacity diverge and change its sign inferring from the relation $C_{R} = \left(\frac{\partial^{2} G_{R}}{\partial T^{2}_{R}}\right)_{P}$. For the non-black hole phase or hot gas phase, the Gibbs free energy is zero. In the viable range of the nonextensive parameter, $\epsilon < \epsilon_G$, there exists the point that the Gibbs free energy of the hot gas phase and one of the black hole phase are equal. At this point, it is possible to obtain the phase transition corresponding to the first-order phase transition called the Hawking-Page phase transition.

\begin{figure}[ht]\centering
\includegraphics[width=7.8cm,height=7.2cm]{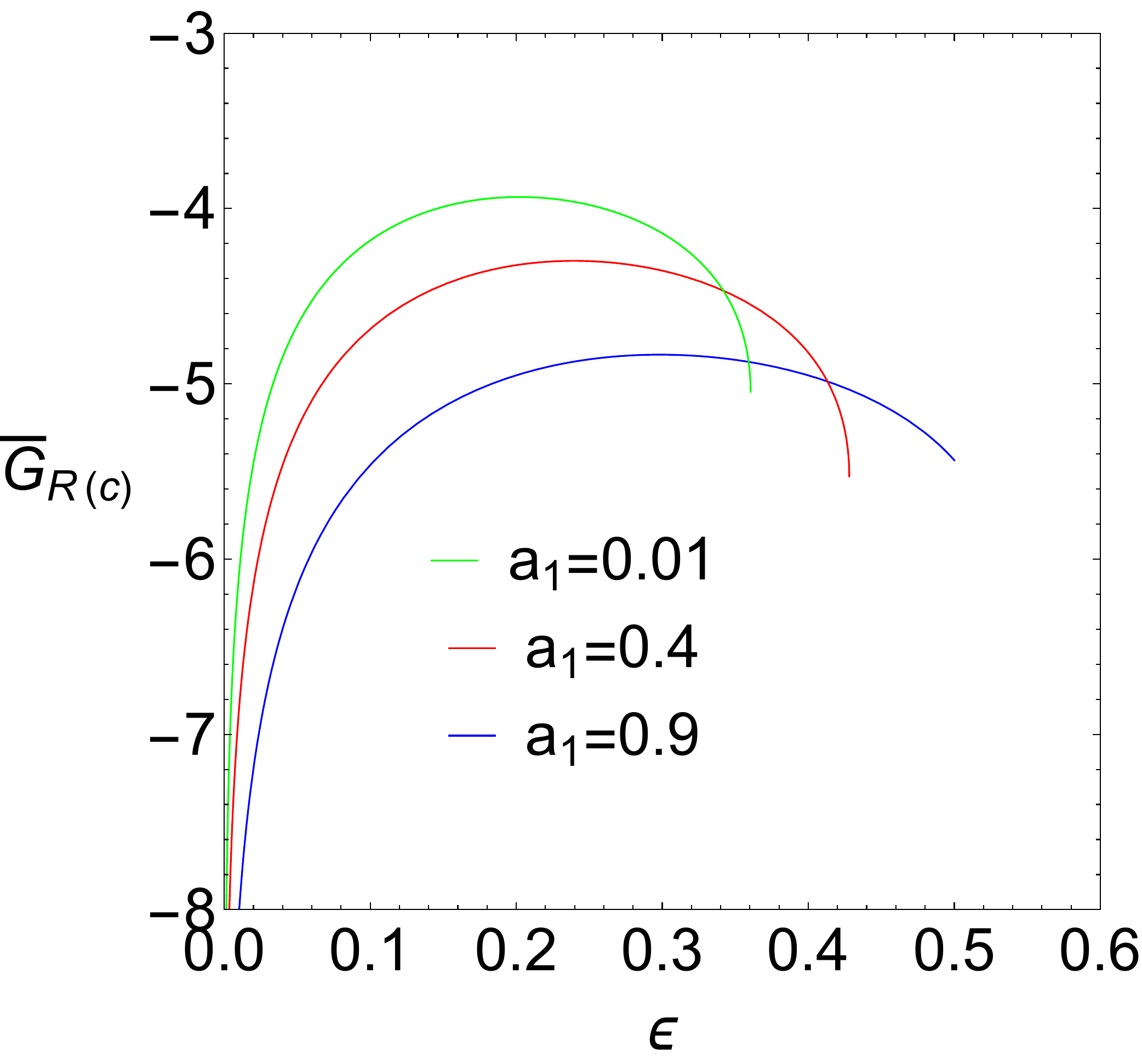}
\quad
\includegraphics[width=7.8cm,height=7.2cm]{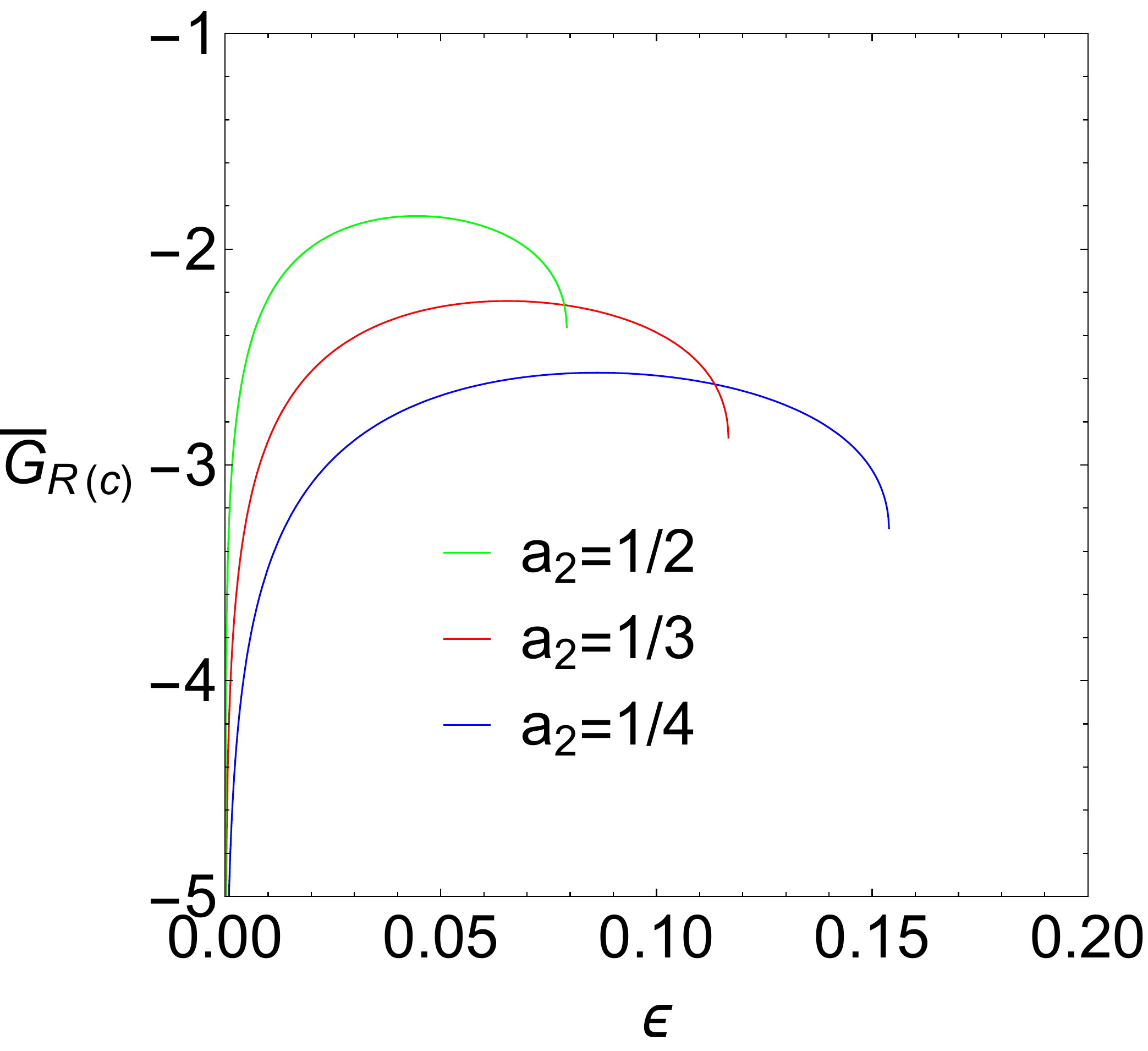}
\caption{The Gibbs free energy for the system evaluated at the cosmological horizon, $\overline{G}_{R(c)}(\epsilon,a_0,a_1,a_2)$ for $0 < \epsilon \leq \epsilon_{G}$ with fixing $a_{0} = 1/2$, $a_2=0.1$ (left) and $a_1=0.1$ (right).}
\label{Gibbs free energy for rc}
\end{figure}

Now, let us study the stability of the system evaluated at the cosmological horizon. It is found that there are no extremum points for $T_{R(c)}$ corresponding to non-cusps in the $\overline{G}_{R(c)}-\overline{T}_{R(c)}$ diagram as shown in the right panel of Fig.~\ref{effective Gibbs}. Using the horizon equations as $f(r_{b}) = 0$ and $f(r_{c})=0$, the mass and pressure can be expressed in terms of $r_b$ and $r_c$ as follows

\begin{eqnarray}
\displaystyle
M &=& \frac{r_{b}r_{c} \big[c_{1} - c_{2}(r_{b} + r_{c})\big]}{2 \big[c_{0} + c_{1} (r_{b} + r_{c}) - c_{2} (r_{b}^{2} + r_{b}r_{c} + r^{2}_{c})\big]}, \label{effective mass} \\
\displaystyle
P &=& -\frac{3}{8\pi}\left[\frac{1}{c_{0} + c_{1} (r_{b} + r_{c}) - c_{2} (r^{2}_{b} + r_{b}r_{c} + r^{2}_{c})}\right]. \label{effective pressure}
\end{eqnarray}
Since we have been interested in the systems which undergo the isobaric process, i.e. $dP=0$, the cosmological horizon radius $r_c$ in terms of the black hole horizon one $r_b$. As a result. the relation in dimensionless variables is expressed as

\begin{equation}
y(x) = \frac{1}{2} \left(a_{1} - x + \sqrt{\frac{4(1+a_{0}) + a_{2}(a_{1}^{2} + 2a_{1}x - 3x^{2})}{a_{2}}}\,\right).
\label{y=y(x)}
\end{equation}
By substituting $y$ corresponding to $x_{+}$ in $G_{R(c)}(y)$, one can obtain $G_{R(c)} = G_{R(c)}(\epsilon,a_{0},a_{1},a_{2})$. However, the expression of $G_{R(c)}(\epsilon,a_{0},a_{1},a_{2})$ is too lengthy, we do not need to show it explicitly here. It is shown numerically that $G_{R(c)}(\epsilon)$ is always negative for $0<\epsilon<\epsilon_G$ as illustrated in Fig.~\ref{Gibbs free energy for rc}. Therefore, the dRGT black hole is globally stable for the nonextensive parameter being in the range $0<\epsilon<\epsilon_G$.

\subsection{Effective system approach} 

In this subsection, the thermodynamic behavior of the black hole is studied by considering the multi-horizon black hole as a single effective system instead of two systems separately defined at $r_b$ and $r_c$. An advantage of this approach is solving the issue of nonequilibrium without imposing that the systems are in the quasi-equilibrium state (as required for the separated system approach). Furthermore, as seen previously, the black hole system can be thermally stable using the R\'{e}nyi statistics. Then, we are interested in the effective system explained by the R\'{e}nyi entropy. The entropy of the effective system is assumed to be the sum of the R\'{e}nyi entropies of both separated systems. Note that this sum obeys the additive composition rule of the R\'{e}nyi entropy. Thus, the entropy of the effective system is given by

\begin{equation}
S=S_{R(b)}+S_{R(c)} 
=\frac{1}{\lambda}\ln\Big[\big(1+\lambda\pi r_{b}^{2}\big)\big(1+\lambda\pi r_{c}^{2}\big)\Big].
\label{total S}
\end{equation}

In order to restrict the well-defined entropy, $\lambda$ is chosen to be positive ($0 < \lambda < 1$). For this effective system approach, the mass; $M = M(S,P)$ is also treated as the enthalpy of the system. The first law of thermodynamics for the effective system is, therefore, written as

\begin{equation}
dM = T_{eff}dS + V_{eff}dP, \label{1st law eff app}
\end{equation}
where $T_{eff}$ and $V_{eff}$ are the effective temperature and the effective volume, respectively. The pressure of this effective system is also defined as the same as one in the separated system approach, i.e., $P=\frac{3}{8\pi}m_{g}^{2}$. The above first law is expected to recover the first laws for separate systems in Eq.~\eqref{1st law sep app} with a suitable limit as will be discussed later.

According to the first law in Eq. \eqref{1st law eff app}, the effective temperature can be computed as follows \cite{Nakarachinda:2021jxd}

\begin{equation}
T_{eff} = \left(\frac{\partial M}{\partial S}\right)_{P} = \frac{\displaystyle\left(\frac{\partial M}{\partial r_{b}}\right)_{r_{c}}\left(\frac{\partial P}{\partial r_{c}}\right)_{r_{b}} -\left(\frac{\partial M}{\partial r_{c}}\right)_{r_{b}}\left(\frac{\partial P}{\partial r_{b}}\right)_{r_{c}}}{\displaystyle\left(\frac{\partial S}{\partial r_{b}}\right)_{r_{c}}\left(\frac{\partial P}{\partial r_{c}}\right)_{r_{b}} +\left(\frac{\partial S}{\partial r_{c}}\right)_{r_{b}}\left(\frac{\partial P}{\partial r_{b}}\right)_{r_{c}}}. 
\label{effTempEq}
\end{equation}

The above expression is indeed obtained from choosing the change of the entropy from Eq. \eqref{total S} as follows

\begin{equation}
dS=dS_{R(b)}-dS_{R(c)}=\left(\frac{\partial S}{\partial r_{b}}\right)_{r_{c}}dr_{b}-\left(\frac{\partial S}{\partial r_{c}}\right)_{r_{b}} dr_{c}.\label{dS eff}
\end{equation}

Note that the negative sign in front of $dS_{R(c)}$ is introduced from the fact that, for an observer who stays between the black hole and cosmological horizons, the direction of heat transfer for the system evaluated at $r_c$ is opposite to that at $r_b$. In other words, when energy transfers from inside to outside of the horizons, the observer experiences positive energy from the black hole horizon but negative energy from the cosmological one. In addition, the effective temperature can be expressed in terms of ones for the separated systems as 

\begin{equation}
\displaystyle
\frac{1}{T_{eff}} = \left(\frac{\partial S}{\partial M}\right)_{P} = \left(\frac{\partial S_{R(b)}}{\partial M}\right)_{P} - \left(\frac{\partial S_{R(c)}}{\partial M}\right)_{P} = \frac{1}{T_{R(b)}} + \frac{1}{T_{R(c)}}.
\label{effTempEq2}
\end{equation}

Interestingly, the definition of effective temperature in Eq. \eqref{effTempEq} can avoid a singularity when $T_{R(b)}=T_{R(c)}$ corresponding to the extremal black hole ($r_{b} = r_{c}$). The usual definition of the change of the entropy, $dS=dS_{R(b)}+dS_{R(c)}$, provides the effective temperature as $T_{eff}=\left(\frac{1}{T_{R(b)}}-\frac{1}{T_{R(c)}}\right)^{-1}$ which obviously diverges for the extremal black hole. Furthermore, the effective temperature can be reduced to the temperature of the separated system evaluated at the black hole horizon for the limit $r_{c} \rightarrow \infty$ and the temperature of the separated system evaluated at the cosmological horizon for the limit $r_{b} \rightarrow 0$;

\begin{eqnarray}
\lim_{r_{c} \rightarrow \infty} T_{eff} &=& T_{R(b)},\label{Teff to Tb}\\
\lim_{r_{b} \rightarrow 0} T_{eff} &=& T_{R(c)}\label{Teff to Tc}.
\end{eqnarray}

These reductions in the effective temperature are also seen from Eq.~\eqref{effTempEq2}, such that $T_{R(c)}$ goes to infinity as $r_c\to\infty$. The effective temperature in the limit $r_c\to\infty$ becomes the temperature for the separated system evaluated at the black hole horizon. It is also found that $T_{R(b)}$ goes to infinity as $r_b\to0$. The effective temperature in the limit $r_b\to0$ becomes the temperature for the separated system evaluated at the cosmological horizon.

Note that in a consideration of the limit $r_c\to0$, the effective temperature still approaches the black hole temperature $\lim_{r_{c} \to0} T_{eff} = T_{R(b)}$. Even though the range of $r_c$ is not valid at zero, this limit is just to eliminate the contribution due to the system evaluated at $r_c$ from the effective system. The interesting point is that this limit is the same one in which the effective volume reduces to volume for the separated system evaluated at $r_b$ as will be discussed later.

It is very important to note that the heat term for the effective system $T_{eff}dS$ can be reduced to those for the separated systems evaluated at $r_b$ and $r_c$, $\pm T_{R(b,c)}dS_{R(b,c)}$. However, if the usual change of entropy is applied, there will be no negative sign in front of the heat term for the system evaluated at $r_c$. This is another advantage point of choosing the change of the entropy as shown in Eq. \eqref{dS eff}. 

\begin{figure}[ht]\centering
\includegraphics[width=8cm,height=8cm]{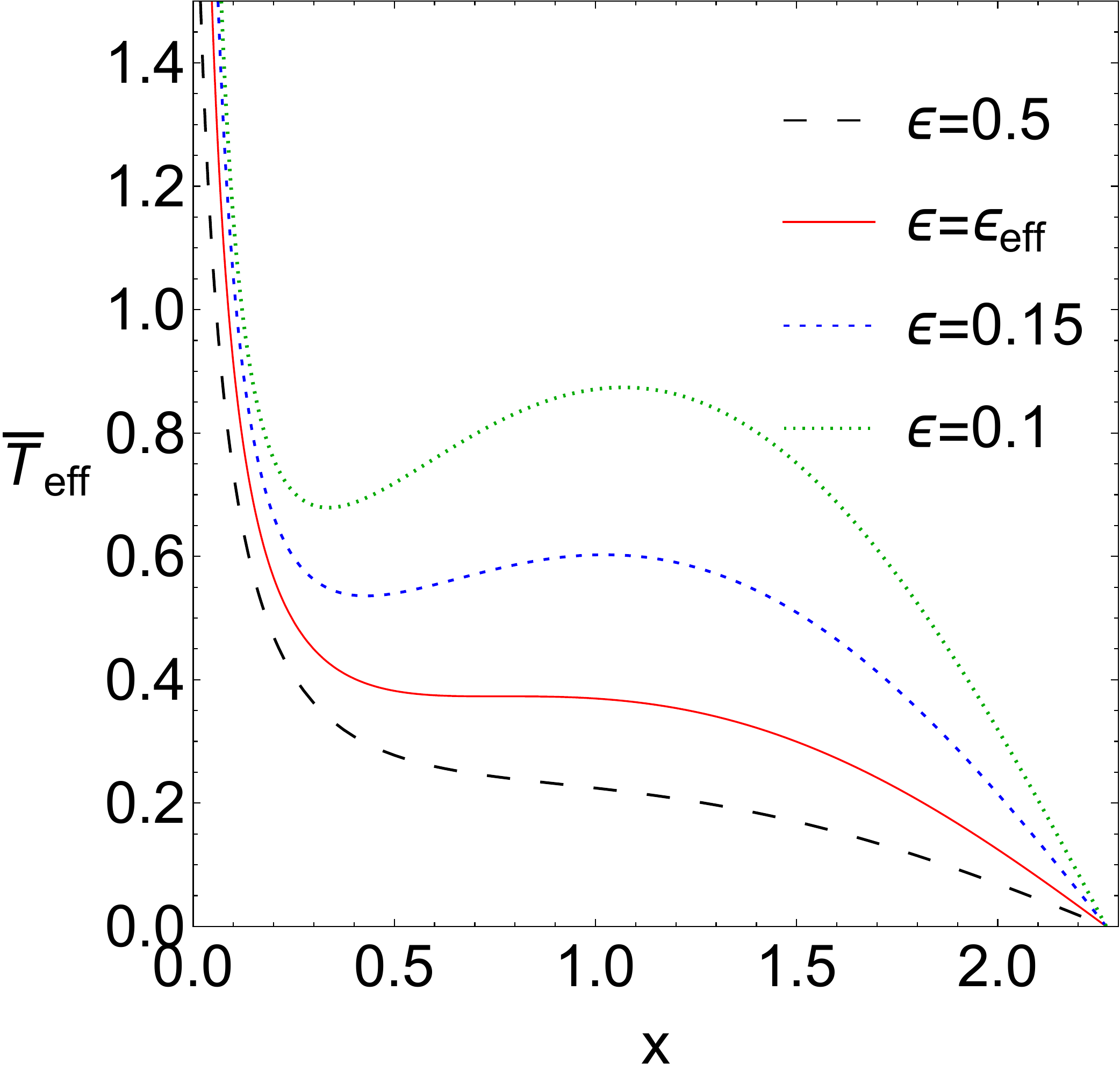}
\quad
\includegraphics[width=8cm,height=8cm]{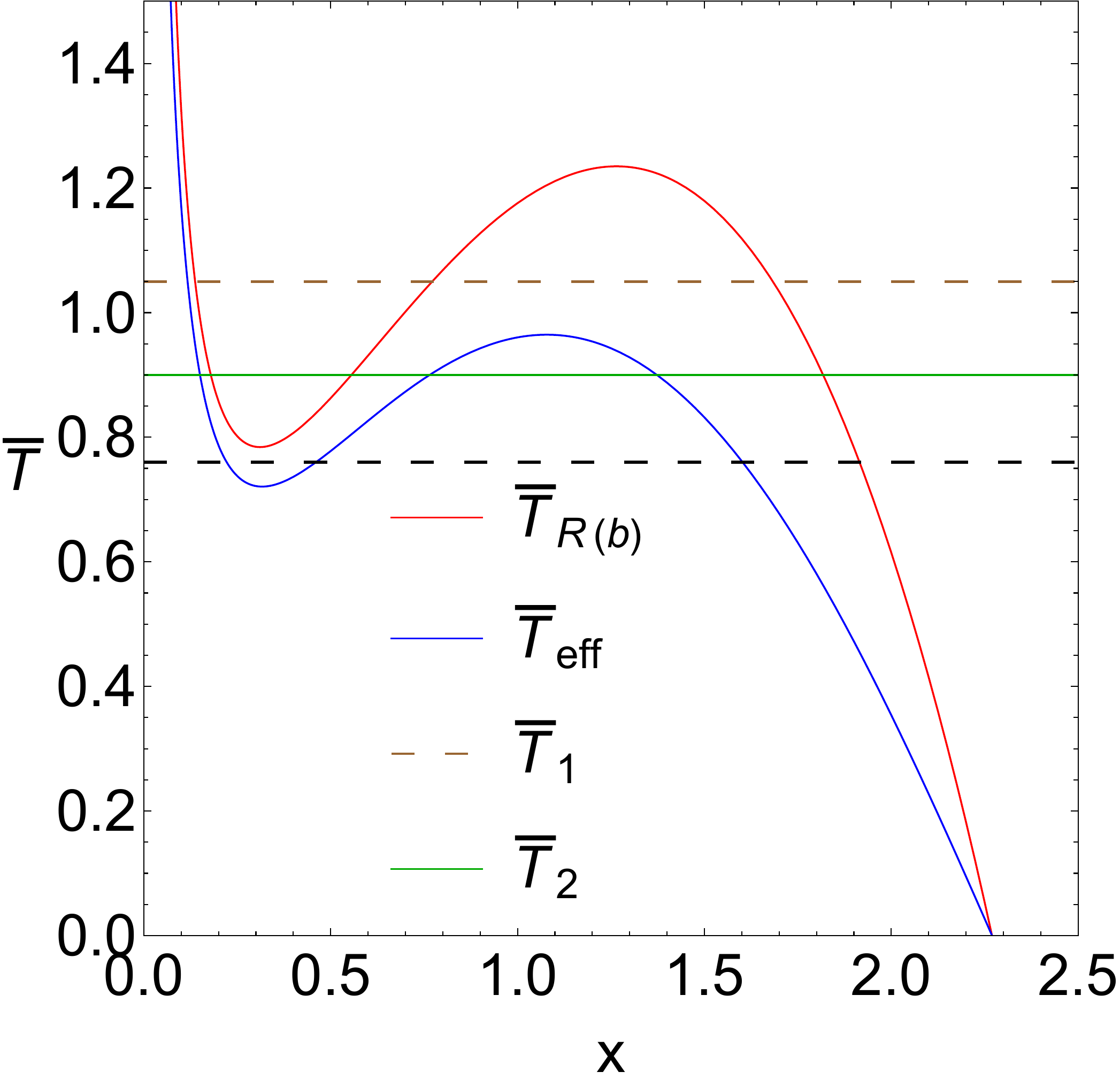}
\caption{Left panel shows the temperature profile of the effective system ($\overline{T}_{eff}=\frac{M}{a_{2}}T_{eff}$) with various values of $\epsilon$ $(\epsilon_{eff} = 0.264388)$ by fixing $a_{0}=1/2$, and $a_{1}=0.1 = a_{2}$. Right panel shows the comparison of temperatures for the separated system evaluated at black hole horizon and the effective system by fixing $a_{0}=1/2$, $a_{1}=0.1 = a_{2}$ and $\epsilon=0.09$.}
\label{effective temperature}
\end{figure}

From Eq.~\eqref{effTempEq}, the effective temperature can be rewritten in terms of only variable $x$ by using the fact that, with fixing the thermodynamic pressure, the cosmological horizon radius depends on the black hole horizon one as shown in Eq.~\eqref{y=y(x)}. It is found that it is too lengthy and not suitable to show here. The behavior of the effective temperature can be illustrated in the left panel of Fig. \ref{effective temperature}. Obviously, the existence of a range with the positive slope in $\overline{T}_{eff}=r_VT_{eff}$ depends on the value of the nonextensive parameter $\epsilon$ similar to $\overline{T}_{R(b)}$. Indeed, the nonextensive parameter $\epsilon$ (or $\lambda$) needs to be sufficiently small (or large) for having a positive slope in $\overline{T}_{eff}$. Therefore, the nonextensivity in the R\'enyi entropy is still required in order to form the locally stable black hole in the effective system approach. In other words, the black hole is always locally unstable using the Gibbs-Boltzmann statistics ($\lambda\to0$ or $\epsilon\to\infty$). When the black hole is in its extremal limit, the effective temperature goes to zero. This value can be seen by rearranging the expression in Eq.~\eqref{effTempEq2} as $T_{eff}=\frac{T_{R(b)}T_{R(c)}}{T_{R(b)}+T_{R(c)}}$. Although $T_{R(b)}$ and $T_{R(c)}$ go to zero for the extremal black hole, the effective temperature is finite (equal to zero) because the numerator approaches zero faster than the denominator.

The comparison of temperature profiles between $\overline{T}_{R(b)}$ and $\overline{T}_{eff}$ can be shown in the right panel of Fig. \ref{effective temperature}. It is seen that $\overline{T}_{R(b)}$ does not much deviate from $\overline{T}_{eff}$ for small-sized black holes while the difference gets large for moderated- and large-sized black holes. At a high temperature such as $\overline{T}_{1}$ in the right panel of Fig. \ref{effective temperature}, the black hole in the separated system approach is locally stable, but one in the effective system approach is locally unstable. On the other hand, at a low temperature such as $\overline{T}_{3}$, the black hole in the effective system approach is locally stable, while one in the separated system approach is locally unstable. Finally, at an intermediate temperature such as $\overline{T}_{2}$, both black holes in the separated and effective system approaches can be locally stable. If the temperature and the radius of the black hole can be observed, it is able to distinguish which approach prefers, since the black hole in the effective system approach is always larger than one in the separated system approach. This may be useful if one wants to assess the possibility of which approach is preferred over the other, one may need access to observations on the radii of the black holes along with their temperatures.

Let us consider the local stability condition on the nonextensive parameter. In order to find the bound on $\epsilon$, one can use the same strategy as done in the previous subsection by considering the extremum points of the temperature. Since the effective temperature depends on both $x$ and $y$, one has to use the fact that $y$ can be expressed in terms of $x$ with keeping the pressure constant as shown in Eq.~\eqref{y=y(x)}. Eventually, the condition of finding extremum points of the effective temperature with respect only to the variable $x$ is given by

\begin{eqnarray}
F_{eff}(x,y)
\equiv\left(\frac{\overline{T}_{R(b)}+\overline{T}_{R(c)}}{\overline{T}_{R(c)}}\right)^2\frac{\partial \overline{T}_{eff}}{\partial x}
=F_b+Y(x,y)=0,\label{cond for local inst}
\end{eqnarray}
where $F_b$ has been previously defined in Eq.~\eqref{extrema of temperature b} and

\begin{eqnarray}
Y(x,y) 
=\frac{dy}{dx}\frac{d\overline{T}_{R(c)}}{dy}\frac{\overline{T}^{2}_{R(b)}}{\overline{T}^{2}_{R(c)}}.
\end{eqnarray}

The function $F_b$ is the concave function of $x$ as we have already mentioned. Since $dy/dx<0$ ($y$ increases or decreases as $x$ decreasing or increasing, respectively) and $d\overline{T}_{R(c)}/dy>0$ ($T_{R(c)}$ is the increasing function of $y$ as illustrated in the right panel in Fig.~\ref{temperature of separated}), the function $Y(x,y)$ always has negative value. As a result, the function $F_{eff}(x,y)$ is the concave function which is lower than $F_b$. This is why the behaviors of $T_{R(b)}$ and $T_{eff}$ are similar.

Using Eq.~(\ref{y=y(x)}), the function $Y(x,y)$ in Eq.~\eqref{cond for local inst} can be written in terms of only $x$. Generally, one can solve Eq.~\eqref{cond for local inst} for two positive real roots of $x$. These two values of $x$ correspond to two extrema of the effective temperature. As seen in Fig.~\ref{effective temperature}, a range of the radius of the locally stable black hole (the black hole with a positive slope of temperature) lies between the two extrema. Hence, a critical point on $\epsilon$, in which the locally stable black hole phase appears or disappears, can be evaluated by merging the two extrema of $T_{eff}$ as a single point. In other words, this critical point can be obtained when the maximum of $F_{eff}$ yields $F_{eff}=0$ itself. Therefore, for the existence of a locally stable phase, it is possible to find an upper bound of the nonextensive parameter denoted as $\epsilon_{eff}$, i.e., $\epsilon\leq\epsilon_{eff}$ is the local stability condition for the effective system approach. This bound should be expressed in terms of the dRGT parameters; $a_0, a_1,$ and $a_2$. Unfortunately, the expression of $\epsilon_{eff}=\epsilon_{eff}(a_0,a_1,a_2)$ is very complicated and it is then difficult to analyze on the nonextensive length scale as done for the separated system. We have used the numerical method to find it. It is also expected that the value of $\epsilon_{eff}$ with arbitrary $a_0, a_1$ and $a_2$ should smaller than the value of $\epsilon_C$, since $F_{eff}$ is always lower than $F_b$ as we have analyzed previously. It is interestingly found that, when $a_{1}$ and $a_{2}$ are small, $\epsilon_{eff}$ is approximated by scaling $\epsilon_C$ as

\begin{equation}
\epsilon_{eff(app)} \approx 0.70616\,\epsilon_C,
\label{ep_eff app}
\end{equation} 

\begin{figure}[ht]\centering
\includegraphics[width=8cm,height=8cm]{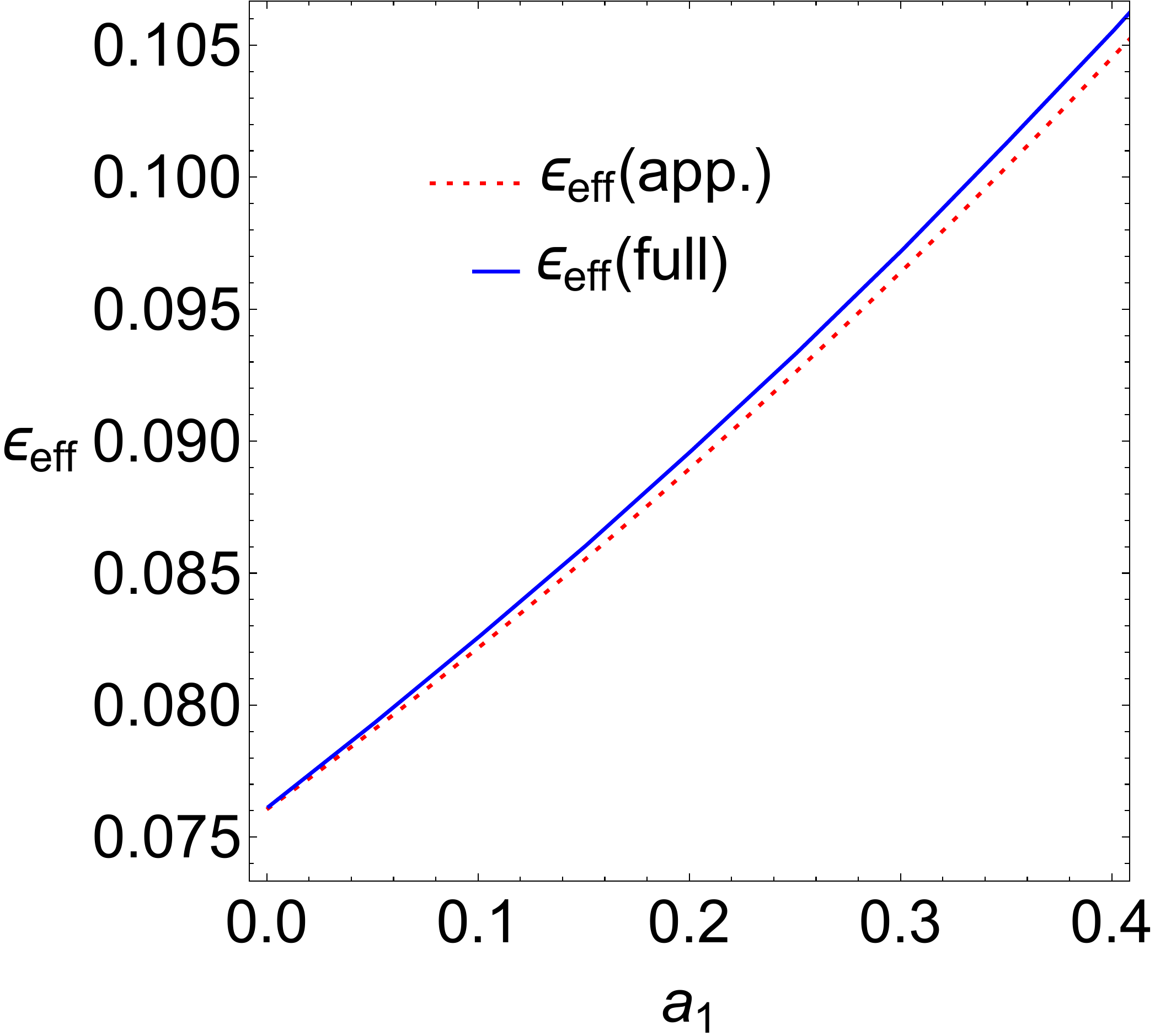}
\quad
\includegraphics[width=8.2cm,height=7.91cm]{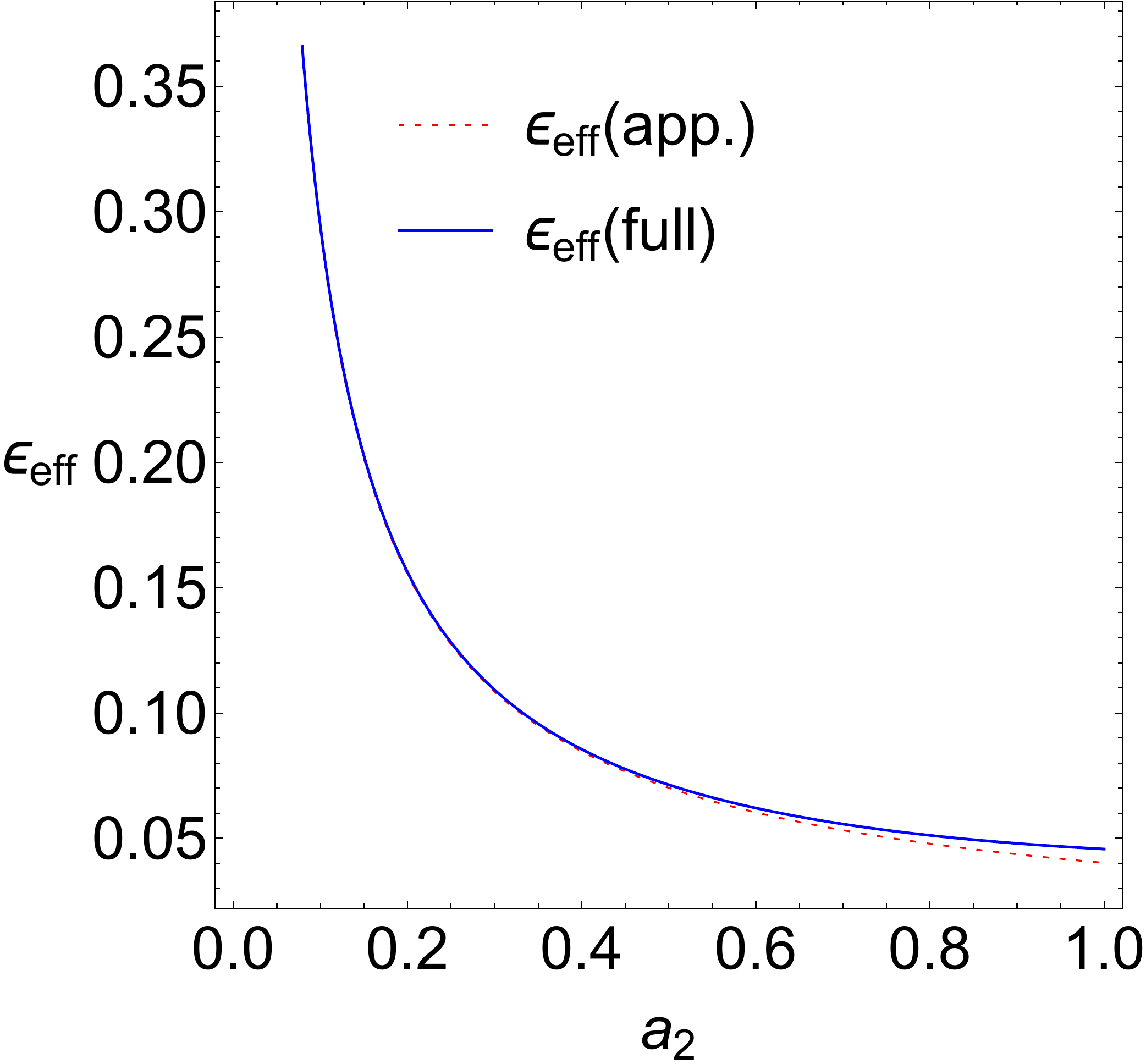}
\caption{Left (Right) panel shows the comparison between $\epsilon_{eff}$ obtained from the full data and approximation versus $a_{1}$ ($a_2$) by fixing $a_{0}=1/2$ and $a_{2}=1/3$ ($a_1=1/3$).}
\label{effective a1}
\end{figure}

The coefficient is actually the same ratio of $\epsilon_{eff(dS)}/\epsilon_{C(dS)}$ for the Sch-dS black hole as investigated in Refs.~\cite{Tannukij:2020njz,Nakarachinda:2021jxd}. Fig.~\ref{effective a1} shows that the approximated bound $\epsilon_{eff(app)}$ is very closed to the exact bound $\epsilon_{eff(full)}$. Since $\epsilon_{eff(app)}$ is slightly smaller than $\epsilon_{eff(full)}$, this guarantees that the black hole with $\epsilon\leq\epsilon_{eff(app)}$ is indeed locally stable.

Moreover, it is very important to note that the (upper) bound on $\epsilon$ for the local stability in the effective system approach is stronger than that in the separated system approach, $\epsilon_{eff}<\epsilon_C$. One can conclude that the effective system approach requires more nonextensivity (more deviates from the Gibbs-Boltzmann statistics) than the separated one in order to obtain the locally stable dRGT black hole. 

\begin{figure}[ht]\centering
\includegraphics[width=9.2cm,height=8cm]{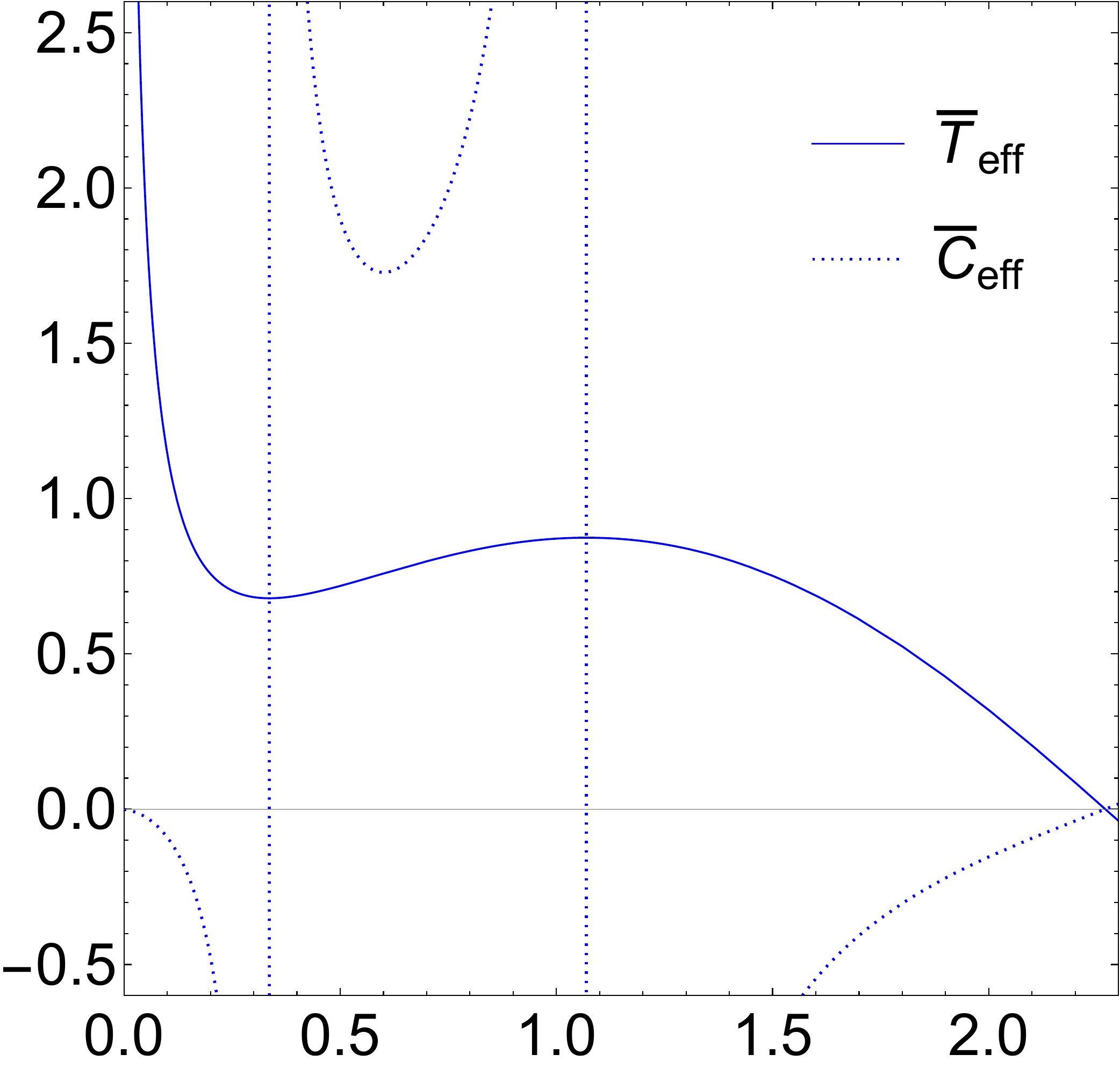}
\caption{The effective temperature and heat capacity profiles with respect to $x$ for fixing $a_{0}=1/2$, $a_{1}=0.1 = a_{2}$ and $\epsilon=0.1$.}
\label{The plot of effective heat capacity}
\end{figure}

Now, the effective heat capacity at the constant pressure is defined as

\begin{equation}
C_{eff} = \left(\frac{\partial M}{\partial T_{eff}}\right)_{P}.
\end{equation}
The explicit expression of the effective heat capacity is too lengthy, it is not shown here. Similar to the analysis in the previous subsection, the black hole mass $M$ is the monotonically increasing function in $x$ for suitable values of parameters. The sign of the effective heat capacity $C_{eff}$ is directly referred to as the sign of the slope of the effective temperature $T_{eff}$. The heat capacity then diverges at the extremum points of the effective temperature. This feature of dimensionless heat capacity $\overline{C}_{eff}=C_{eff}/r_V$ and temperature $\overline{T}_{eff}$ are shown in Fig.~\ref{The plot of effective heat capacity}. Let us emphasize that the positive effective heat capacity corresponds to the positive slope of the effective temperature. The effective system is locally stable for the moderate-sized black hole on $\epsilon \leq \epsilon_{eff}$.

The thermodynamic volume of the effective system is defined accordingly. The effective volume is computed from \cite{Nakarachinda:2021jxd}

\begin{equation}
\displaystyle
V_{eff} = \left(\frac{\partial M}{\partial P}\right)_{S} = \frac{\left(\displaystyle\frac{\partial M}{\partial r_b}\right)_{r_c}\left(\displaystyle\frac{\partial S}{\partial r_c}\right)_{r_b} +\left(\displaystyle\frac{\partial M}{\partial r_c}\right)_{r_b} \left(\displaystyle\frac{\partial S}{\partial r_b}\right)_{r_c}}{\left(\displaystyle\frac{\partial P}{\partial r_b}\right)_{r_c} \left(\displaystyle\frac{\partial S}{\partial r_c}\right)_{r_b} +\left(\displaystyle\frac{\partial P}{\partial r_c}\right)_{r_b}\left(\displaystyle\frac{\partial S}{\partial r_b}\right)_{r_c}}.
\label{effective volume}
\end{equation}

Note that this effective volume can be reduced to the volumes of each system in the separated system point of view as follows:

\begin{eqnarray}
\lim_{r_{c} \rightarrow 0} V_{eff} &=& V_{b},\\
\lim_{r_{b} \rightarrow 0} V_{eff} &=& V_{c}.
\end{eqnarray}

In the limits, $r_c\to0$ and $r_b\to0$, the first law for the effective system approach \eqref{1st law eff app} recovers the first laws for the separated system evaluated at the black hole and cosmological horizons \eqref{1st law sep app}, respectively. Similarly to the previous analysis, it must be emphasized that although the limit $r_c\to0$ is not valid, such a limit eliminates contributions from $r_c$ just like it did in the temperature case. Furthermore, the effective volume in Eq. \eqref{effective volume} can be written in terms of $V_{b}$ and $V_{c}$ as \cite{Sriling:2021lpr,Nakarachinda:2021jxd}

\begin{equation}
\displaystyle
V_{eff} = T_{eff} \left(\frac{V_{b}}{T_{R(b)}} + \frac{V_{c}}{T_{R(c)}}\right).
\end{equation}

From the above expression, the effective volume is obviously positive for the viable range of the dRGT parameters $a_{0}$, $a_{1}$ and $a_{2}$ yielding positive $V_b$ and $V_c$. Note also that, using the usual definition of the change of the entropy, $dS=dS_{R(b)}+dS_{R(c)}$, the effective volume, $V_{eff}=T_{eff}\left(\frac{V_{b}}{T_{R(b)}}-\frac{V_{c}}{T_{R(c)}}\right)$, is possible to be negative which is unphysical. As seen in Eq.~\eqref{Vbc}, $V_b$ and $V_c$ are taken in the same functions of $r_b$ and $r_c$, respectively. $V_c$ is always greater than $V_b$ because $r_c>r_b$. Using this fact and Eq.~\eqref{effTempEq2}, it is straightforwardly obtained that the effective volume is always greater than the volume of the separated system evaluated at the black hole horizon. It is important to emphasize that the (effective) volume and pressure can be concurrently positive for the dRGT black hole described by the effective system approach.

Let us consider the global stability by using the effective Gibbs free energy. The effective Gibbs free energy is defined as

\begin{equation}
G_{eff} = M - T_{eff}S.
\end{equation}

\begin{figure}[ht]\centering
\includegraphics[width=8cm,height=8cm]{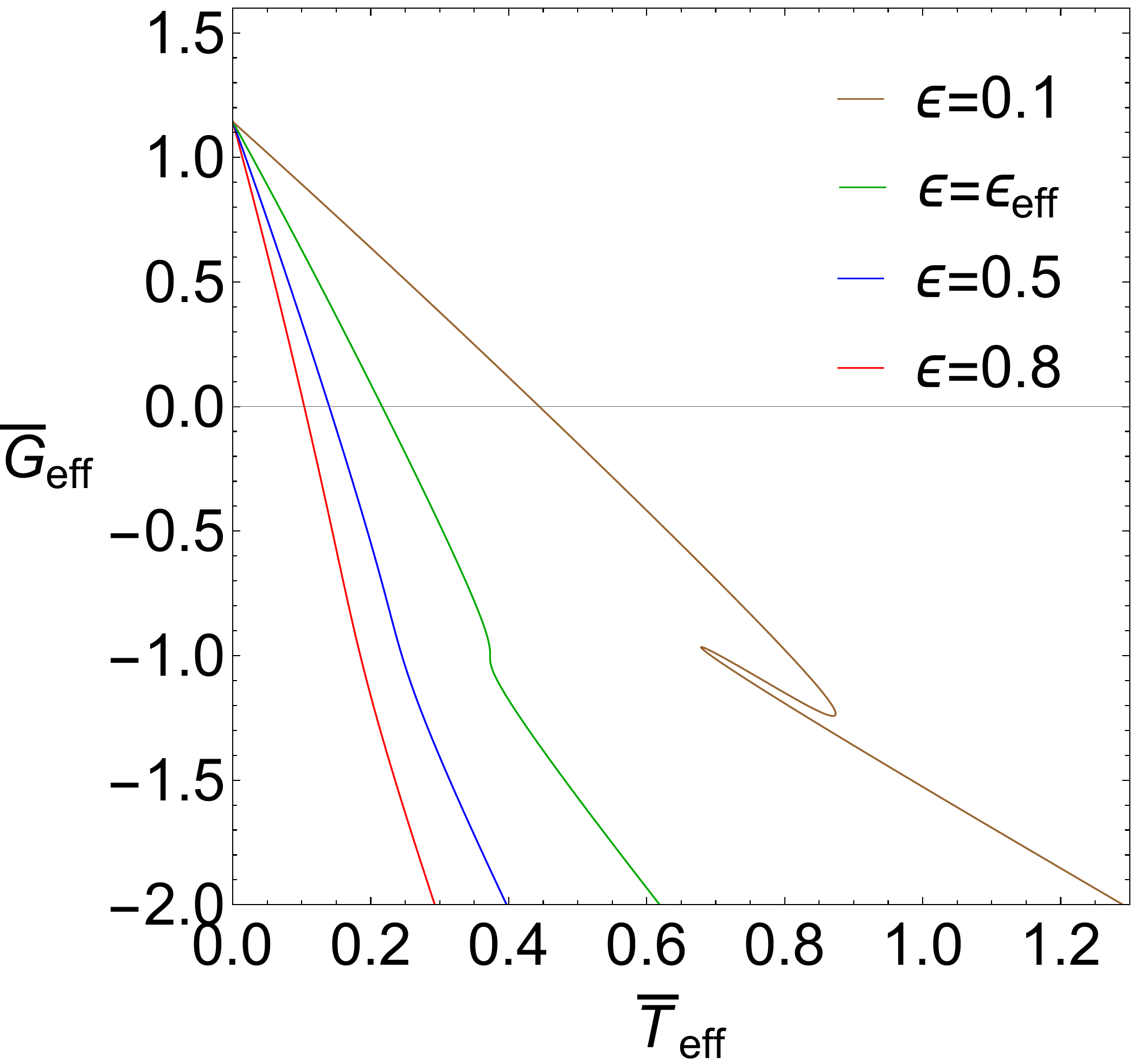}
\quad\quad
\includegraphics[width=8cm,height=8cm]{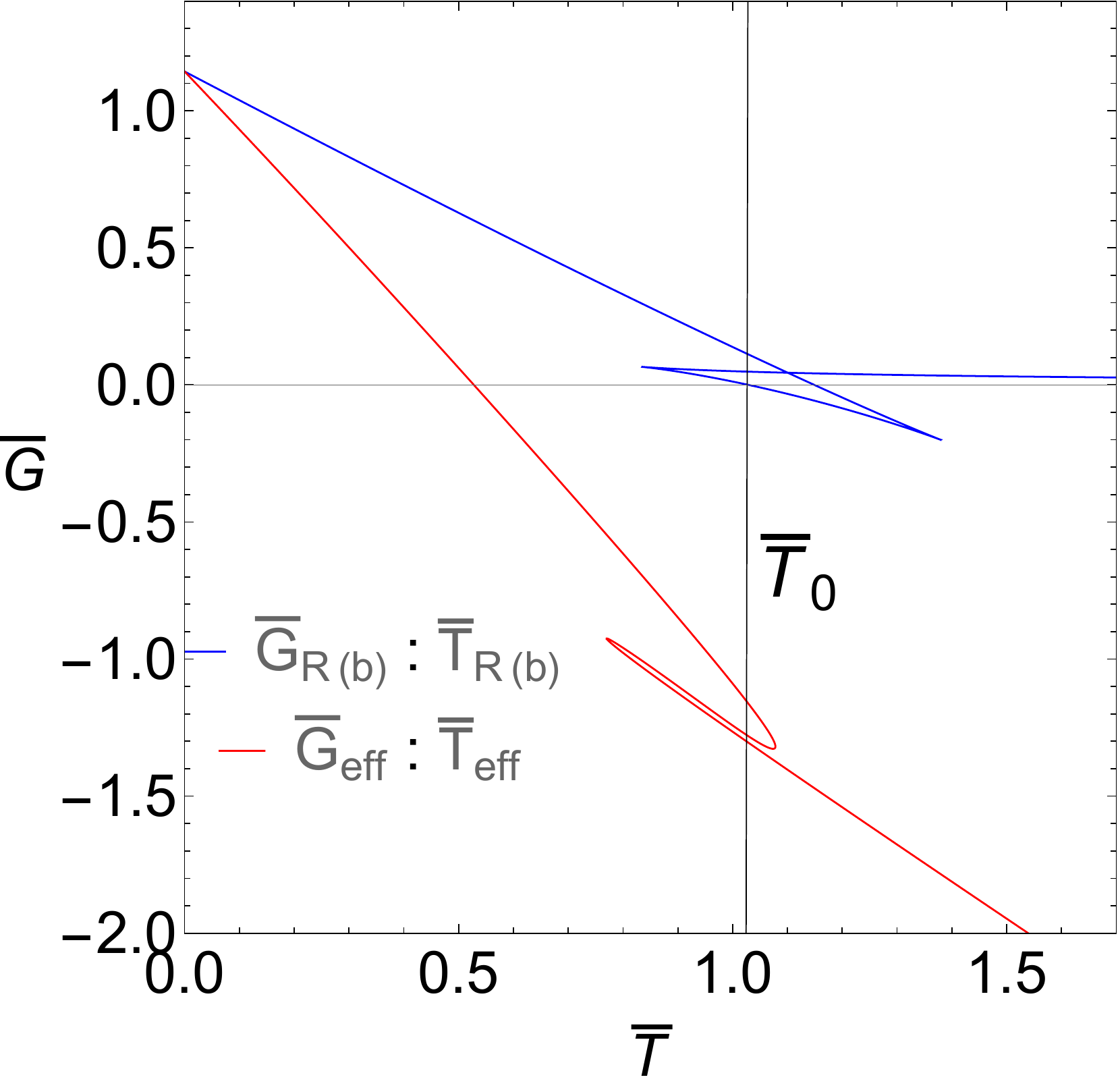}
\caption{Left panel shows the profile of the effective Gibbs free energy versus effective temperature with various values of $\epsilon$ $(\epsilon_{eff}=0.264388)$. Right panel shows the profiles of the Gibbs free energy versus the temperature for the separated system evaluated at the black hole horizon (blue line) and effective system (red line) with fixing $\epsilon=0.08$. In this figure, we have used $a_{0}=1/2$ and $a_{1}=0.1 = a_{2}$.}
\label{The plot of effective Gibbs free energy}
\end{figure}

It can be written in terms of the dimensionless variables as $\overline{G}_{eff}(x,y)=r_V G_{eff}$. Instead of explicitly showing its full expression, the behavior of the effective Gibbs free energy with the various values of $\epsilon$ is illustrated in the left panel of Fig.~\ref{The plot of effective Gibbs free energy}. It is seen that when $\epsilon<\epsilon_{eff}$, there exist cusps that approximately correspond to the phase transitions between the locally stable-unstable black hole phase. 

Furthermore, one can notice that the cusps in the effective system approach are not peaked as those in the separated system approach are (see the right panel of Fig.~\ref{The plot of effective Gibbs free energy}). It is because, due to the modification of $dS$ in Eq.~\eqref{dS eff}, the change of $G_{eff}$ is not taken in the usual form, $dG=-SdT+VdP$, but is written as $dG_{eff}=-SdT_{eff}+V_{eff}dP+2T_{eff}dS_{R(c)}$. Then, the first and second derivatives of the effective Gibbs free energy with respect to effective temperature are not exactly equal to $-S$ and $-C_{eff}/T_{eff}$, respectively. However, the cusps in the $\overline{G}_{eff}-\overline{T}_{eff}$ diagram are closed to the extremum points of $\overline{T}_{eff}$ or divergent points of $\overline{C}_{eff}$ as seen in Fig.~\ref{The plot of effective Gibbs free energy}. It can be estimated that the left/right cusp is around the local minimum/maximum point of $\overline{T}_{eff}$. The range of the black hole being locally stable (moderate-sized black hole) lies between these cusps. From the left panel of Fig.~\ref{The plot of effective Gibbs free energy}, one also sees that the locally stable black hole approximately has the lowest Gibbs free energy when it is at the right cusp. Therefore, the bound on $\epsilon$ for the moderate-sized black hole being globally stable can be estimated from the Gibbs free energy at the right cusp being zero. By numerical analysis, it is found that the moderate-sized black hole always has negative Gibbs free energy. Apart from the local stability condition, $\epsilon\leq\epsilon_{eff}$, there is no further bound on $\epsilon$ for the global stability of the black hole in the effective system approach. In other words, the locally stable black hole described by the effective system approach is always globally stable. Using the result of Eq.~\eqref{ep_eff app}, the nonextensive length scale can be obtained as

\begin{eqnarray}
\frac{r_{h}}{L_\lambda}
\sim \sqrt{0.70616}\,a_1\sqrt{a_2}\approx0.84a_1\sqrt{a_2}.
\end{eqnarray}
Emphasize that this expression is applicable only for the case of $a_1$ and $a_2$ being small. 

Additionally, the global stability bound for the black hole described by the separated system approach is stronger than the bound for the black hole described by the effective system approach, e.g., $\epsilon_G\approx0.456\epsilon_C$ for setting $a_0=1/2$ and $a_1=a_2=0.1$. Hence, the nonextensivity in the effective system approach is less required than that in the separated one in order to obtain the locally and globally stable black hole. 

Since the effective Gibbs free energy of the moderate-sized black hole is always negative, the hot gas which has zero Gibbs free energy will form the stable black hole via the Hawking-Page phase transition. Moreover, the effective Gibbs free energy is discontinuous when the phase transition occurs. It implies that the phase transition between the non-black hole and black hole phase is a zeroth-order type. This is an important distinguishable feature between the separated and effective system approaches. For example, from the right panel of Fig.~\ref{The plot of effective Gibbs free energy}, one can see that, at a certain temperature represented as $\overline{T}_0$, the Hawking-Page phase transitions from the hot gas to the stable black hole phases can occur for both separated and effective systems as the first-order type (the slope $d\overline{G}_{eff}/d\overline{T}_{eff}$ jumps) and the zeroth-order type (the value of $\overline{G}_{eff}$ jumps), respectively.

\section{Conclusion}\label{sec:conclusion}

Classically, nothing can escape from the black hole. However, the black hole can emit the Hawking radiation if the effect of quantum mechanics is taken into account. This suggests that the black hole can act as a thermal object. As a result, thermodynamic properties of the black holes have been investigated intensively in order to explore the quantum nature of spacetime. In this work, we investigate the thermodynamic properties of the black hole in de Rham, Gabadadze, and Tolley (dRGT) massive gravity theory based on R\'enyi entropy.

Since the entropy of the black hole is proportional to its area, the black hole entropy is supposed to be a nonextensive quantity. Therefore, the thermodynamics of the black hole should be based on nonextensive statistics. R\'enyi entropy is one of the entropies which can characterize the nonextensive nature of the thermodynamic system. Consequently, it is possible to obtain the thermodynamically stable Schwarzschild (Sch) and Schwarzschild-de Sitter (Sch-dS) black holes in the context of R\'enyi entropy while they are unstable if the entropy of the black hole is classified as the usual Gibbs-Boltzmann (GB) entropy \cite{Nakarachinda:2021jxd,Czinner:2015eyk,Tannukij:2020njz}. In this work, we explore how nonextensivity can influence the thermodynamic stability of the black hole in the dRGT massive gravity theory, called the dRGT black hole.      

The dRGT massive gravity theory can provide the asymptotically de Sitter (dS) solution which is compatible with the late-time expansion of the universe. For a spherically symmetric solution, the dRGT black hole can provide corrections to Sch-dS black hole. Therefore, we analyze how the thermodynamic properties of the dRGT black hole are modified compared to the Sch-dS black hole. For the Sch-dS black hole, it is well known that either the thermodynamic pressure or volume is negative. However, in the dRGT case, it is possible to define the positive pressure  by keeping a volume positive. This is one of the worthy properties of the dRGT black hole compared to the Sch-dS black hole. One of the signatures of the black hole with asymptotically dS spacetime is that there exist two horizons between which we live.
Therefore, there are corresponding thermodynamic systems with generally different temperatures. In order to deal with this kind of black hole, we classified our analysis into two categories: the separated system approach and the effective system approach. 

For the separated system approach, the systems are assumed to be far from each other enough and are not significantly different in temperature. By adopting the first law of black hole thermodynamics, we examine the nonextensivity by replacing the GB entropy with the R\'enyi entropy. The pressure is defined to be proportional to the graviton mass in the same fashion as that $P \sim \Lambda$ in the Sch-dS black hole. With this definition, we found that the pressure is positive with some range of the graviton mass parameters by keeping volume positive as shown in Fig.~\ref{The volume of dRGT}. Moreover, we analyze the behavior of temperature and heat capacity of both separated systems with defining the temperature properly via the first law.  We find that the dRGT black hole can be locally stable due to the presence of nonextensivity with the dimensionless nonextensive parameter $\epsilon=1/(\lambda\pi r_V^2)$, where $r_V$ is Vainshtein radius, less than an upper bound $\epsilon_C$ following Eq.~\eqref{condition of local}. Without graviton mass corrections, the bound of $\epsilon$ reduces to one for the Sch-dS black hole, $\epsilon_{C(dS)}$. It implies that the nonextensive length scale $L_\lambda$ must be less than the Hubble radius $L_\lambda \lesssim 0.268\, H_0^{-1}$, where $H_0$ is Hubble parameter at the present $(H_0\sim m_g\sim L_{mg}^{-1})$. On the other hand, the global stability is investigated by analyzing the behavior of the Gibbs free energy. We find that the global bound of the nonextensive parameter $\epsilon_G$ is stronger than one for the local bound,  $\epsilon < \epsilon_G \approx 0.456\, \epsilon_C$. Remarkably, the transition from the thermal radiation or hot gas phase to the stable black hole phase, so-called Hawking-Page phase transition, of this system is found to be of the first-order type. It is shown in the left panel of Fig.~\ref{effective Gibbs}.

For the effective system approach, the thermodynamic systems are assumed to be described by effective thermodynamic quantities. By following the first law in the same form as one in the separated system approach, the effective quantities are defined by using the criterion such that the heat flow for the system evaluated at the cosmological horizon has the opposite direction to one at the black hole horizon \cite{Nakarachinda:2021jxd}. This comes from the fact that the observer stays between the black hole horizon and the cosmological horizon. We find that, with the (positive) pressure defined in the same way as one in separated system approach, the effective volume is still positive. By using the same strategy as performed in the separated system approach, the bound to obtain the locally stable black hole is found to be stronger than one in the separated system approach as $\epsilon < \epsilon_{eff} = 0.70616 \epsilon_C$. This implies that the thermodynamic stability of the black hole in the effective system approach requires the nonextensivity of the system greater than one in the separated system approach. Surprisingly, this relation gives exactly the same as one for the Sch-dS case even though the effects of graviton mass are included. Moreover, it is found that there exist particular temperatures in which the black hole in both approaches will be locally stable. In this case, the black hole radius in the effective system approach is always larger than one in the separated system approach. Furthermore, there exist particular temperatures for which the black hole is locally stable either in the effective or separated system approach. As a result, these particular temperatures can be used to distinguish between the two approaches. For the global stability, we find that the Gibbs free energy in the range with local stability is always negative. Therefore, the locally stable black hole is always globally stable without another requirement as found in the separated system approach. Interestingly, the Hawking-Page phase transition is found to undergo from hot gas to the black hole with the zeroth-order phase transition.

From our results, the nonextensive bounds get modified due to the additional contribution of graviton mass parameterized by $a_1$ and $a_0$ corresponding to the nonlinear scales above the Vainshtein radius. Interestingly, the correction term due to the graviton mass is scaled by $r_{h}/L_\lambda$ where $r_{h}$ is the horizon. Therefore, for large nonextensivity limit, the R\'enyi entropy can be expressed in the form of $S_R \sim \ln(r_h/L_\lambda)$ which coincides with one for the entanglement entropy $S_e \sim \ln(\xi/a)$ \cite{Vidal:2002rm,Calabrese:2004eu}. Note that $\xi$ and $a$ denote the correlation length and lattice spacing, respectively. As a result, we may argue that the black hole horizon can play the role of the correlation length.
It should be emphasized that this speculation on the relation between the black hole horizon and the correlation length is based on the existence of $r_h/L_\lambda$ which arises from the existence of the graviton mass. This may shed light on the interplay between the nature of entanglement and gravitational interaction contributed by the graviton mass.

It is important to note that in the dRGT massive gravity theory, there exists the cutoff scale $r_{\Lambda_3} \sim (m_g^2 M_{Pl})^{-1/3} = (M_{Pl}/M)^{1/3} r_V$ which is much smaller than the Vainshtein radius. The dRGT black hole with a radius comparable to such the cutoff is not trustable. However, at this cutoff scale, the effect of graviton mass is strongly suppressed and then the gravitational interaction should be understood through general relativity. With this respect, our analysis is not enough to demonstrate correspondences between entropy which may be related to the microscopic states of quanta spacetime and the graviton mass. In particular, the dRGT black hole can be treated as a classical background spacetime while the radiation from the black hole can be treated as a quantum effect without being influenced by graviton mass.

It is noteworthy to emphasize here that, for our analysis, we collaborate the R\'enyi entropy with the black hole thermodynamics by adopting the first law of thermodynamics derived from the gravitational description with GB statistics. While there might be other ways to collaborate the R\'enyi entropy to the black hole thermodynamics \cite{Nojiri:2022aof,Nojiri:2022sfd}, this prescription allows us to define the proper thermodynamic quantities based on the thermodynamic laws. It would be interesting to investigate the first law from the gravitational description with the R\'enyi entropy. We leave this investigation for further works.

\section*{Acknowledgement}

This research project is supported by National Research Council of Thailand (NRCT): NRCT5-RGJ63009-110. PW is supported by National Science, Research and Innovation Fund (NSRF) through grant no. R2565B030.

\bibliography{ref.bib}

\end{document}